\pdfoutput=1
\documentclass[a4paper,12pt]{article}

\usepackage{jheppub} 

\usepackage{tangocolors}
\usepackage{mathrsfs}
\usepackage{color}

\hypersetup{ linktoc=all,
    colorlinks, linkcolor={palatinateblue},
    citecolor={brightpink}, urlcolor={amaranth}
}

\def\sideremark#1{\ifvmode\leavevmode\fi\vadjust{\vbox to0pt{\vss
 \hbox to 0pt{\hskip\hsize\hskip1em
 \vbox{\hsize2cm\tiny\raggedright\pretolerance10000
 \noindent #1\hfill}\hss}\vbox to8pt{\vfil}\vss}}}%
\DeclareSymbolFont{extraup}{U}{zavm}{m}{n}
\DeclareMathSymbol{\varheart}{\mathalpha}{extraup}{86}
\DeclareMathSymbol{\vardiamond}{\mathalpha}{extraup}{87}
\makeatletter
\renewcommand*{\@fnsymbol}[1]{\ensuremath{\ifcase#1\or \clubsuit \or \vardiamond \or \varheart\or
    \spadesuit\or \mathparagraph\or \|\or **\or \dagger\dagger
    \or \ddagger\ddagger \else\@ctrerr\fi}}
\makeatother

\definecolor{rosy}{RGB}{230,235,252}
\definecolor{myframetitle}{RGB}{90,89,170}
\definecolor{myblocktitle}{RGB}{140,185,249}
\definecolor{mytitle}{RGB}{10,80,26}

\definecolor{darkgreen}{RGB}{27,130,45}
\definecolor{darkblue}{rgb}{0,0,0.3}
\definecolor{darkred}{rgb}{0.7,0,0}

\definecolor{light gray}{RGB}{220,220,220}
\definecolor{dark purple}{RGB}{108,0,217}
\definecolor{pink}{RGB}{190,20,100}
\definecolor{orang}{RGB}{193,63,0}
\definecolor{green}{RGB}{11,98,17}
\definecolor{darkpink}{RGB}{153,0,76}
\definecolor{bluegreen}{RGB}{0,102,102}
\definecolor{greenlagan}{RGB}{0,102,0}
\definecolor{redgreen}{RGB}{102,102,0}
\definecolor{Redgreen}{RGB}{153,76,0}
\definecolor{vividviolet}{rgb}{0.62, 0.0, 1.0}
\definecolor{amaranth}{rgb}{0.9, 0.17, 0.31}
\definecolor{palatinateblue}{rgb}{0.15, 0.23, 0.89}
\definecolor{brightpink}{rgb}{1.0, 0.0, 0.5}
\definecolor{cornflowerblue}{rgb}{0.39, 0.58, 0.93}
\definecolor{deepcarminepink}{rgb}{0.94, 0.19, 0.22}
\definecolor{radicalred}{rgb}{1.0, 0.21, 0.37}

\DeclareFontFamily{OT1}{rsfs}{}

\DeclareFontShape{OT1}{rsfs}{m}{n}{ <-7> rsfs5 <7-10> rsfs7 <10->rsfs10}{}

\DeclareMathAlphabet{\mycal}{OT1}{rsfs}{m}{n}

\makeatletter \@addtoreset{equation}{section}

\begin{document}



\title{Transient Acceleration after Non-minimal M-flation Preheating}

\author[]{Amjad Ashoorioon,}
\author[]{Kazem Rezazadeh}
\affiliation[]{\small School of Physics, Institute for Research in Fundamental Sciences (IPM),\\
P.O. Box 19395-5531, Tehran, Iran}

\emailAdd{amjad@ipm.ir}
\emailAdd{kazem.rezazadeh@ipm.ir}


\abstract{Light massive preheat fields acquire a non-vanishing dispersion during parametric resonance from their quantum particle production. This in turn will modify the inflaton potential, which in some cases can induce a transient period of acceleration. We illustrate this phenomenon in the setup of non-supersymmetric non-minimal M-flation (non-$\mathbb{M}$-flation) which has some motivations from the brane compactifications in string theory. Implementing a lattice simulation by the LATTICEEASY code, we compute the potential correction term in our scenario and show that the modified term indeed causes the Universe to make a transition from the decelerated expansion to a temporary phase of acceleration. The correction term reduces to some extent the number density of the particles generated during preheating, but the efficiency of preheating remains still enough to have successful particle production after inflation. We also compute the spectrum of the gravitational waves (GWs) generated during preheating in our setup by using the LATTICEEASY code. Although the peak frequency remains almost the same, the inclusion of the correction term reduces the amplitude of the gravitational spectrum by almost one order of magnitude.}

\subheader{IPM/P-2023/05}


\maketitle


\section{Introduction}
\label{section:introduction}

How the cold, empty Universe after inflation has evolved to our observable Universe has been the subject of rigorous investigation in the past. Aside from reheating \cite{Linde:1990flp}, where the inflaton decays perturbatively to other forms of matter, it is now known that coupling of another field to the inflaton, while it oscillates around its minimum, can lead to a burst of particle production. This is known as parametric resonance or stochastic preheating \cite{Traschen:1990sw, Shtanov:1994ce, Kofman:1994rk, Kofman:1997yn}. During this period, the number density of the preheat fields can grow or decay due to the expansion of the Universe, although it mainly grows most of the time. The products of this process finally interact with each other and thermalize at a temperature known as reheating temperature. In most inflationary models, even the ones that have roots in string theory, the existence and coupling of such preheat fields are assumed in an ad-hoc manner.

An inflationary model which has well-based theoretical motivations from string theory is Matrix inflation, or simply M-flation \cite{Ashoorioon:2009wa, Ashoorioon:2011ki}. Different aspects of this model were investigated in the literature \cite{Ashoorioon:2009sr, Ashoorioon:2013oha, Ashoorioon:2014jja}. In this class of models, we deal with three $N\times N$ Hermitian matrices $\Phi_{i}$ ($i=1,2,3$), and the potential is assumed to be a function of $\Phi_{i}$ or the commutations $\left[\Phi_{i},\Phi_{j}\right]$. Therefore, the potential of $\Phi_{i}$ appears in these models as a function of $\mathrm{Tr}\,\Phi_{i}^{2}$, $\mathrm{Tr}\,\epsilon_{ijk}\Phi_{i}\left[\Phi_{j},\Phi_{k}\right]$, and $\mathrm{Tr}\,\left[\Phi_{i},\Phi_{j}\right]^{2}$. In Ref. \cite{Ashoorioon:2009wa, Ashoorioon:2011ki} it was discussed how the potential can be motivated from the brane dynamics in string theory.

The M-flation scenario can resolve the fine-tuning problem of the self-couplings in chaotic inflation models \cite{Linde:1983gd}, as it has been shown explicitly in \cite{Ashoorioon:2009wa}. In the setting of standard chaotic inflation, in order to make the simple chaotic inflation models compatible with the CMB observations, the potential parameters are required to be very small which is not acceptable from the point of view of quantum field theory. In addition, the inflaton field displacement in the chaotic setups is super-Planckian which is not favored by the quantum field theory considerations, and it also is not consistent with the swampland distance conjecture \cite{Ooguri:2018wrx, Agrawal:2018own}. The M-flation scenario, however, can resolve these problems by providing the potential parameters of order unity, and also presenting sub-Planckian field displacements during inflation \cite{Ashoorioon:2009wa}. In this setup, like other multi-field inflation models, the isocurvature perturbations are present along with the adiabatic one. Of course, due to their masses, their amplitudes decay exponentially by the end of inflation. Since the coupling of these isocurvature fields is related to the self-couplings of the inflaton, their amplitude can be understood exactly. Also at the end of inflation, when the slow-roll condition is violated, these isocurvature fields can act as preheat fields and drain energy from the inflation. Knowing their coupling to the inflaton, one can have exact predictions for the gravitational wave spectrum produced during preheating \cite{Ashoorioon:2013oha}.

Despite these achievements, M-flation suffers from several shortcomings. For instance, although this model is based on string theory, so far the issue of compactification to four dimensions has not been worked out. More strictly, thus far, it is not clear how the 4-dimensional gravity arises when one compactifies the extra dimensions. In addition, in order to solve the problems of the smallness of the potential self-couplings from the bare ones, the ranks of the matrices, which represent the number of D3-branes in the stack, are required to be of order $N\sim10^{4}-10^{5}$, which is large and makes the backreaction on the underlying background geometry an important issue that needs to be addressed. It is also not compatible with the latest bound on the tensor-to-scalar ratio by the Planck data \cite{Planck:2018jri}.

In the setup of non-minimal M-flation (non-$\mathbb{M}$-flation) \cite{Ashoorioon:2019kcy}, in which the inflaton field couples non-minimally to gravity, we can resolve some of these problems. The non-minimal coupling to gravity is something that has been observed in the renormalization of the gravity in presence of matter. In fact, it has been shown that the loop correction to the graviton-scalar-scalar vertex generates a term proportional to $\ensuremath{\xi\Lambda^{2}\phi^{2}/M_P^{2}}$ with $\xi\simeq \mathcal{O}(1)$ \cite{Ashoorioon:2011aa}, where $\Lambda$ is the cutoff of the theory. Noting that $\Lambda$ is at most $M_P$, the resulting coupling is of order one at most, and cannot justify the large values of $\xi$ if it is needed. In the $\mathbb{K}$L$\mathbb{M}$T inflation model proposed by Kachru, Kallosh, Linde, Maldacena, McAllister, and Trivedi \cite{Kachru:2003sx}, where inflation is driven by the moduli corresponding to the separation of D3 $\mathrm{\bar{D}}3$,  when the superpotential is proportional to the volume modulus, it was possible to obtain large non-minimal coupling parameters \cite{Ashoorioon:2019kcy}. One should note that the idea of the non-minimal coupling to the gravitational field has been adopted in the Higgs inflation model \cite{Bezrukov:2007ep}. It will be more convenient to analyze the model, if we go through a conformal transformation to the Einstein frame. In non-$\mathbb{M}$-flation, it is possible to improve the observational consistency of the symmetry-breaking potential with the Planck 2018 CMB data \cite{Planck:2018jri} compared to the minimal M-flation predictions \cite{Ashoorioon:2019kcy}. The number of the D3-branes in non-$\mathbb{M}$-flation is obtained to be of order $N\lesssim10^{2}$ which is considerably less than the required number in the minimal M-flation setup, which is one appealing consequence of non-minimal coupling in the model \cite{Ashoorioon:2019kcy}. Another appealing feature of the non-$\mathbb{M}$-flation, in the supersymmetric setup, was the hope that preheating around the symmetry-breaking vacuum could be addressed. However as we will revisit this problem here, we notice that the problem persists in the non-minimal symmetry-breaking M-flation. That is why we turn our attention to a supersymmetric setup in this paper.

As noted above, after inflation, the spectator fields may play an important role in the Universe's evolution. In particular, their coupling to the inflaton field can provide a successful mechanism of parametric resonance that explains the particle production in the post-inflationary Universe. This process is also known as the preheating process during which the Universe experiences a decelerating expansion \cite{Kofman:1994rk, Kofman:1997yn}. During preheating, the number of particles can increase significantly. The particles produced at this stage are far away from thermal equilibrium and have enormously large occupation numbers. In the next stage, the interactions of the existing fields with the other fields become important, and the previously produced particles decay. In the final step, the cosmic fluid will be thermalized, and this process generally takes a very long time which is inversely proportional to the coupling constants.

The quantum dispersion of the preheat field squared leads to the modification of the potential which changes the evolution of the post-inflationary Universe. The phenomenon was mainly noticed in the symmetry-breaking potentials where the quantum fluctuations of the preheat field restore the symmetry and lead to the production of topological defects and secondary stage of inflation \cite{Kofman:1995fi, Felder:2000sf}. The period of accelerated expansion which takes place after preheating may have significant theoretical and observational implications. In particular, such a transient period of acceleration could be crucial in solving the moduli and gravitino problems. This phase of acceleration can be regarded as secondary non-thermal inflation and it may be an alternative to the thermal inflation \cite{Lyth:1995hj} proposed for solving these problems \cite{Felder:2000sf}. In the example that we consider here, a similar phenomenon is noted for a light massive preheat field that acquires a dispersion during inflation preheating. The mass term acts for a period of time like a cosmological constant that lifts the inflaton potential and leads to accelerated expansion. Another manifestation of particle production during inflation is prolonging the length of inflation or even making the inflation happen on a steep potential \cite{Adshead:2018oaa, Adshead:2012kp}.

Preheating after inflation exponentially increases the number density of particles in some specific momentum instability bands. This corresponds to inhomogeneities that source gravitational waves at the second order in perturbation theory \cite{Khlebnikov:1997di, Easther:2006vd, Dufaux:2007pt}. Such inhomogeneities source the tensor perturbations that lead to stochastic gravitational wave background (SGWB). The SGWB signals from the preheating were first investigated by \cite{Khlebnikov:1997di}. In \cite{Easther:2006vd}, the authors applied numerical computations to solve the equations of the metric perturbations in the Fourier space and evaluated the GW spectrum from preheating in inflationary models with energy scales much lower than the GUT scale. The GWs from preheating also has been studied in the context of a variety of inflationary settings, including gauge preheating after axion inflation \cite{Adshead:2018doq, Adshead:2019lbr, Cui:2021are}, self-resonance after single-field inflation, oscillon formation \cite{Lozanov:2019ylm, Antusch:2016con, Amin:2018xfe, Hiramatsu:2020obh, Kou:2021bij}, and tachyonic preheating from a waterfall transition \cite{Garcia-Bellido:2007nns, Garcia-Bellido:2007fiu, Dufaux:2010cf}. Recently, it has been shown that the spectrum of the stochastic background of the gravitational wave is different from those produced by a first-order phase transition \cite{Ashoorioon:2022raz}. Unfortunately, the current knowledge based on existing literature implies that the GW signal from preheating is generally beyond the reach of foreseeable known GW experiments: either the frequency of the resultant GW signal is too high to be covered, or the amplitude of the signal is too weak to be detectable with near future GW detectors. However, plans to develop tabletop experiments based on resonant cavity experiments that are immersed in constant electromagnetic fields, have been put forward \cite{Berlin:2021txa}.


In \cite{Easther:2006vd}, the GWs from preheating have been investigated in a simple inflationary setup with the quadratic potential $V(\phi)=\mu^{2}\phi^{2}/2$ where $\mu$ is a constant parameter with dimensions of mass. In their investigation, they have shown that for different values of the inflaton effective mass in the wide range $\mu = 10^{-18}-10^{-6}m_{P}$ where $m_P$ is the Planck mass, the amplitude of the GWs is almost identical and it is of order $\Omega_{\mathrm{GW}}h^{2}\sim10^{-11}-10^{-10}$, but the spectrum appears in different frequencies. For the important case of $\mu=10^{-6}m_{P}$ the spectrum appears in the frequencies $f\sim10^{8}-10^{10}\mathrm{Hz}$. In this paper, we refer to this case as the benchmark reheating scenario, and we will compare our findings with the results of that model\footnote{In our recent paper \cite{Ashoorioon:2022raz}, it was shown that in the lower energy scales, the amplitude of the GWs spectrum shows some dependency on the parameter $\mu$.}.

We study the effect of the preheating process on the inflaton potential in the framework of non-$\mathbb{M}$-flation, but away from the symmetry-breaking configuration. The potential thus has only one minimum. We use the numerical code LATTICEEASY \cite{Felder:2000hq} to compute the root mean square (r.m.s.) of the scalar spectator field in our model, and then we use it to calculate the modified term in the potential. We include the contribution of the additional potential energy in the ensuing background dynamics. Depending on the potential parameters and the properties of the dominant preheat field, there will be different preheating scenarios in our setup. We show that in a specific case, the lifted potential is capable of providing a transition to a temporary accelerating phase of expansion. In this paper, we show that this transient period of accelerated expansion suppresses the number density of the produced preheat particles and hence reduces the amplitude of the gravitational waves generated during preheating.

This paper is structured as follows. First, in Sec. \ref{section:model}, we review briefly the basics of the non-$\mathbb{M}$-flation model. Next, in Sec. \ref{section:preheating}, we present the basic equations for the study of preheating in this setup. Subsequently in Sec. \ref{section:transient_acceleration}, we will consider a special case in the setting of non-$\mathbb{M}$-flation that is capable of providing a transient acceleration after preheating. We will then investigate the effect of the transient acceleration on the number density of the produced mediate particles. Then, in Sec. \ref{section:GWs}, we examine the impact of the potential correction term on the spectrum of the stochastic GWs background generated during preheating. Finally, we summarize our findings in Sec. \ref{section:conclusions}.


\section{Non-$\mathbb{M}$-flation}
\label{section:model}

In non-$\mathbb{M}$-flation, it is assumed that the matrix fields $\Phi_{i}$ couple non-minimally to the gravitational field, so the action of the scalar fields of the model in the Jordan frame can be written as
\begin{equation}
\label{SJ-Phi}
S_{J}=\int d^{4}x\sqrt{-g}\left[\frac{1}{2}\left(1+\xi\sum\limits _{i=1}^{3}\mathrm{Tr}\left(\Phi_{i}^{2}\right)\right)R-\frac{1}{2}\sum\limits _{i=1}^{3}{\rm Tr}\left(\partial_{\mu}\Phi_{i}\partial^{\mu}\Phi_{i}\right)-V\left(\Phi_{i},\left[\Phi_{i},\Phi_{j}\right]\right)\right] \; .
\end{equation}
Throughout this paper, we work in the units where the reduced Planck mass is set to unity, $M_{P}\equiv1/\sqrt{8\pi G}=1$. The coupling strength of the matrix fields with gravitation is determined by the coupling constant $\xi$. We do not include the gauge fields in the present analysis. The inflationary potential $V$ in this action is a function of the matrices $\Phi_{i}$, and their commutations $\left[\Phi_{i},\Phi_{j}\right]$. In Ref. \cite{Ashoorioon:2009wa} it has been discussed that it is possible to derive the potential from the examination of the string theory branes. In that reference, it has also shown that in the special context of string theory and to the dominant order of $\Phi_{i}$ and $\left[\Phi_{i},\Phi_{j}\right]$, the potential takes the following form
\begin{equation}
\label{V-Phii}
V\left(\Phi_{i},\left[\Phi_{i},\Phi_{j}\right]\right)=\mathrm{Tr}\left(-\frac{\lambda}{4}\left[\Phi_{i},\Phi_{j}\right]\left[\Phi_{i},\Phi_{j}\right]+\frac{i\kappa}{3}\epsilon_{jkl}\left[\Phi_{k},\Phi_{l}\right]\Phi_{j}+\frac{m^{2}}{2}\Phi_{i}^{2}\right) \; ,
\end{equation}
where $\lambda$ is a dimensionless constant, while the constants $\kappa$ and $m$ have dimensions of mass. These constants have some interpretations from string theory: the constant $\lambda=8\pi g_{s}=2g_{YM}^{2}$ is related to the string coupling constant $g_s$ or the Yang-Mills constant $g_{YM}$, the constant $\kappa=\hat{\kappa}g_{s}\sqrt{8\pi g_{s}}$ has relation with the Ramond-Ramond antisymmetric form strength $\hat{\kappa}$, and the constant $m$ comes from the three spatial coordinates along the D3-branes in the metric of the background SUGRA theory \cite{Ashoorioon:2009wa, Ashoorioon:2011ki}. The background is a solution to SUGRA equations with a constant dilaton if $\lambda m^{2}=4\kappa^{2}/9$. However, in this paper, we consider more general configurations and deviate from this condition.

Since the matrices $\Phi_{i}$s are $N-$dimensional, we deal with $3N^{2}$ real scalar fields, and hence the study of the model in its most general form becomes very complicated. Instead, we introduce the matrices $J_{i}$ ($i=1,2,3$) that are the $N\times N$ generators of the SU(2) algebra, and satisfy the commutation relation $\left[J_{i},J_{j}\right]=i\,\epsilon_{ijk}\,J_{k}$. Thus, we can decompose the inflaton matrices into two parts,
\begin{equation}
\label{Phii-phihat}
\Phi_{i}=\hat{\phi}J_{i}+\Psi_{i} \, .
\end{equation}
The first and second parts in this decomposition are respectively the parallel and perpendicular components to the $N\times N$ matrix representation of the SU(2) algebra (i.e. $\mathrm{Tr}\left(J_{i}\Psi_{i}\right)=0$). In Refs. \cite{Ashoorioon:2009wa, Ashoorioon:2013oha}, it has been argued that if the fields $\Psi_{i}$ are turned off at the early stages, they remain classically zero during the inflation. Thus we call them the spectator fields. Ignoring the spectator fields, the inflationary trajectory is determined by the $\hat{\phi}$ field which is the component of the matrices along the SU(2) generator. Accordingly, the action \eqref{SJ-Phi} can be rewritten as
\begin{equation}
\label{SJ-phihat}
S_{J}=\int d^{4}x\sqrt{-g}\left[\frac{1}{2}\left(1+\frac{\xi}{M_{P}^{2}}\mathrm{Tr}\left(J_{i}^{2}\right)\hat{\phi}^{2}\right)R+\mathrm{Tr}\left(J_{i}^{2}\right)\left(\frac{1}{2}\left(\frac{d\hat{\phi}}{dt}\right)^{2}-\frac{\lambda}{2}\hat{\phi}^{4}+\frac{2\kappa}{3}\hat{\phi}^{3}-\frac{m^{2}}{2}\hat{\phi}^{2}\right)\right] \, ,
\end{equation}
with ${\rm Tr}\left(J_{i}^{2}\right)=N\left(N^{2}-1\right)/4$. Since the kinetic term in this action is non-canonical, we redefine the scalar field as
\begin{equation}
\label{phi-phihat}
\phi=\sqrt{\mathrm{Tr}\left(J_{i}^{2}\right)}\,\hat{\phi} \, .
\end{equation}
As a result, the action takes the following form
\begin{equation}
\label{SJ}
S_{J}=\int d^{4}x\sqrt{-g}\left[\frac{1}{2}\left(1+\xi\phi^{2}\right)R+\frac{1}{2}\left(\frac{d\phi}{dt}\right)^{2}-V_{0}(\phi)\right] \, .
\end{equation}
Introducing the effective coupling constants $\lambda_{\mathrm{eff}}\equiv2\lambda/\mathrm{Tr}\left(J_{i}^{2}\right)$ and $\kappa_{\mathrm{eff}}\equiv\kappa/\sqrt{\mathrm{Tr}\left(J_{i}^{2}\right)}$, the inflaton potential \eqref{V-Phii} can be written as
\begin{equation}
\label{V0}
V_{0}(\phi)=\frac{\lambda_{\mathrm{eff}}}{4}\phi^{4}-\frac{2\kappa_{\mathrm{eff}}}{3}\phi^{3}+\frac{m^{2}}{2}\phi^{2} \, .
\end{equation}
To get rid of the non-minimal coupling to gravity, following Ref. \cite{Bezrukov:2007ep}, we do the following conformal transformation from the Jordan to the Einstein frame,
\begin{equation}
\label{gtildemunu}
\tilde{g}_{\mu\nu}=\Omega^{2}g_{\mu\nu}\, , \qquad \Omega^{2}=1+\xi\phi^{2} \, .
\end{equation}
The tilde represents the quantities in the Einstein frame. This conformal transformation makes the kinetic term in the Jordan frame to be non-canonical, and to resolve this problem, we introduce the new scalar field $\chi$ which is the scalar field in the Einstein frame, and it has a relation to the Jordan frame scalar field $\phi$ as
\begin{equation}
\label{dchidphi}
\frac{d\chi}{d\phi}=\frac{\sqrt{\Omega^{2}+6\xi^{2}\phi^{2}}}{\Omega^{2}}=\frac{\sqrt{\xi\phi^{2}(6\xi+1)+1}}{\xi\phi^{2}+1} \, .
\end{equation}
Therefore, the action in the Einstein frame turns into
\begin{equation}
\label{SE}
S_{E}=\int d^{4}x\sqrt{-\tilde{g}}\left[\frac{1}{2}\tilde{R}+\frac{1}{2}\left(\frac{d\chi}{dt}\right)^{2}-U(\chi)\right] \, ,
\end{equation}
where $U(\chi)$ is the inflationary potential in the Einstein frame and for it we have
\begin{equation}
\label{U}
U(\chi)=\frac{V_{0}\left(\phi(\chi)\right)}{\Omega^{4}\left(\phi(\chi)\right)} \, .
\end{equation}
From the ordinary differential equation \eqref{dchidphi}, the field $\chi$ can be expressed in terms of the field $\phi$ as
\begin{equation}
\label{chi-phi}
\chi=f(\phi)=\sqrt{\frac{1}{\xi}+6}\,\sinh^{-1}\left[\sqrt{\xi\left(6\xi+1\right)}\,\phi\right]-\sqrt{6}\,\tanh^{-1}\left[\frac{\sqrt{6}\,\xi\text{\ensuremath{\phi}}}{\sqrt{\xi(6\xi+1)\text{\ensuremath{\phi}}^{2}+1}}\right] \, .
\end{equation}
It is not simple to solve the above equation to express $\phi$ in terms of $\chi$, and for this purpose, we can use the inverse function,
\begin{equation}
\label{phi-chi}
\phi=f^{-1}(\chi) \, .
\end{equation}
In the numerical calculations, it is more convenient to use the following function which provides an appropriate approximation for $\phi$ versus $\chi$
\begin{equation}
\label{phi-chi-approx}
\phi\approx\frac{a_{1}\chi+a_{2}\chi^{2}+a_{3}\chi^{3}+a_{4}\chi^{4}+a_{5}\chi^{5}+a_{6}\chi^{6}+a_{7}\chi^{7}+a_{8}\chi^{8}+a_{9}\chi^{9}+a_{10}\chi^{10}}{1+b_{1}\chi+b_{2}\chi^{2}+b_{3}\chi^{3}+b_{4}\chi^{4}+b_{5}\chi^{5}+b_{6}\chi^{6}+b_{7}\chi^{7}+b_{8}\chi^{8}+b_{9}\chi^{9}+b_{10}\chi^{10}} \, .
\end{equation}
The numerical coefficients in this equation can be determined by curve-fitting, for each set of model parameters.


\section{Preheating in non-$\mathbb{M}$-flation}
\label{section:preheating}

In this section, we aim to study preheating in the framework of non-$\mathbb{M}$-flation. For this purpose, we take the spectator scalar fields into account, and then the Jordan frame action will be
\begin{equation}
\label{SJ-Psi}
S_{J}=\int d^{4}x\sqrt{-g}\left[\frac{1}{2}\left(1+\xi\phi^{2}\right)R-\frac{1}{2}g^{\mu\nu}\partial_{\mu}\phi\partial_{\nu}\phi-V_{0}(\phi)-\frac{1}{2}\sum\limits _{i}g^{\mu\nu}\partial_{\mu}\Psi_{i}\partial_{\nu}\Psi_{i}-V_{(2)}\left(\phi,\Psi_{i}\right)\right] \, .
\end{equation}
The potential in this equation has been decomposed into two parts,
\begin{equation}
\label{V}
V=V_{0}\left(\phi\right)+V_{(2)}\left(\phi,\Psi_{i}\right) \, .
\end{equation}
The potential $V_{0}\left(\phi\right)$ has already given by Eq. \eqref{V0}, and for the potential $V_{(2)}\left(\phi,\Psi_{i}\right)$, we have
\begin{equation}
\label{V2}
V_{(2)}\left(\phi,\Psi_{i}\right)=\frac{1}{2}M_{\Psi}^{2}(\phi)\Psi_{i}^{2} \, ,
\end{equation}
where
\begin{equation}
\label{MPsi2}
M_{\Psi}^{2}(\phi)=\frac{1}{2}\lambda_{{\rm eff}}\omega(\omega-1)\phi^{2}+2\kappa_{{\rm eff}}\omega\phi+m^{2} \, .
\end{equation}
The parameter $\omega$ takes integers values, and we can calculate $M_\Psi$ which is the effective mass of the $\Psi_i$ modes. There are two sets of solutions for the effective mass of the scalar spectator fields that are known as $\alpha$-modes and $\beta$-modes. For the $\alpha$-modes we have $\omega=-(j+2)$ where $j$ is an integer in the range $0\leq j\leq N-2$. The degeneracy of each $\alpha_j$-mode is $2j+1$, and so the total number of these modes is equal to $(N-1)^2$ \footnote{There are $N^2-1$ zero modes which are replaced with the gauge modes too. We do not take them into account in this work.}. The $\alpha$-mode with $j = 0$ corresponds to the quantum fluctuations in the SU(2) direction and it is an adiabatic mode. By discarding the contribution of this mode, the total number of degeneracy becomes $(N-1)^2 - 1$. For the $\beta$-modes, we have $\omega = j-1$ with $1\leq j\leq N$. Each $\beta$-mode has $2j+1$ degeneracy, and hence the total degeneracy will be resulted in to be $(N+1)^2 - 1$.

Exploiting the conformal transformation \eqref{gtildemunu}, we can simply show
\begin{equation}
\partial_{\mu}\Psi_{i}\partial^{\mu}\Psi_{i}=g^{\mu\nu}\partial_{\mu}\Psi_{i}\partial_{\nu}\Psi_{i}=\Omega^{2}\tilde{g}^{\mu\nu}\partial_{\mu}\Psi_{i}\partial_{\nu}\Psi_{i} \, .
\end{equation}
Substituting this into Eq. \eqref{SJ-Psi}, we obtain the action in the Einstein frame as
\begin{equation}
\label{SE-Psi}
S_{E}=\int d^{4}x\sqrt{-\tilde{g}}\left[\frac{1}{2}\tilde{R}-\frac{1}{2}\tilde{g}^{\mu\nu}\partial_{\mu}\chi\partial_{\nu}\chi-U(\chi)-\frac{1}{2\Omega^{4}}\sum\limits _{i}g^{\mu\nu}\partial_{\mu}\Psi_{i}\partial_{\nu}\Psi_{i}-\tilde{V}_{(2)}\left(\chi,\Psi_{i}\right)\right] \, .
\end{equation}
The new potential in this action will be
\begin{equation}
\label{V2tilde}
\tilde{V}_{(2)}\left(\chi,\Psi_{i}\right)=\frac{V_{(2)}\left(\chi,\Psi_{i}\right)}{\Omega^{4}\left(\phi(\chi)\right)}=\frac{1}{2}\frac{M_{\Psi_{i}}^{2}\left(\phi(\chi)\right)}{\Omega^{4}\left(\phi(\chi)\right)}\Psi_{i}^{2} \, .
\end{equation}
Imposing the redefinition $\tilde{\Psi}_{i}\equiv\Psi_{i}/\Omega$ and also using the relation $\tilde{g}^{\mu\nu}=g^{\mu\nu}/\Omega^{2}$, action \eqref{SE-Psi} turns into
\begin{equation}
\label{SE-Psitilde}
S_{E}=\int d^{4}x\sqrt{-\tilde{g}}\left[\frac{1}{2}\tilde{R}-\frac{1}{2}\tilde{g}^{\mu\nu}\partial_{\mu}\chi\partial_{\nu}\chi-U(\chi)-\frac{1}{2}\sum\limits _{i}\tilde{g}^{\mu\nu}\partial_{\mu}\tilde{\Psi}_{i}\partial_{\nu}\tilde{\Psi}_{i}-\bar{V}_{(2)}\left(\chi,\tilde{\Psi}_{i}\right)+\frac{\dot{\Omega}}{\Omega}\sum\limits _{i}\tilde{\Psi}_{i}\dot{\tilde{\Psi}}_{i}\right] \, .
\end{equation}
The redefined potential in this equation is given by
\begin{equation}
\label{V2bar}
\bar{V}_{(2)}\left(\chi,\tilde{\Psi}_{i}\right)=\frac{1}{2}M_{\tilde{\Psi}}^{2}\tilde{\Psi}_{i}^{2} \, ,
\end{equation}
where
\begin{equation}
\label{MPsit2}
M_{\tilde{\Psi}}^{2}=\frac{1}{\Omega^{2}}\left(M_{\Psi}^{2}-\dot{\Omega}^{2}\right) \, .
\end{equation}
In the above equation, if the second term in the parentheses is much less than the first term, $\dot{\Omega}^{2}\ll M_{\Psi}^{2}$, we will have
\begin{equation}
\label{MPsit2-approx}
M_{\tilde{\Psi}}^{2}\approx\frac{M_{\Psi}^{2}}{\Omega^{2}} \, .
\end{equation}
The use of this approximation will simplify our calculations considerably, in particular for the lattice simulations. We will use this approximation in the case we analyze, and then in the subsequent steps, we verify that it is indeed valid.

The mode $\tilde{\Psi}_{i}(t,\mathbf{x})$ can be decomposed into its Fourier components as follows
\begin{equation}
\label{Psiti-Psitik}
\tilde{\Psi}_{i}(t,\mathbf{x})=\int\frac{d^{3}k}{(2\pi)^{3/2}}\left[\tilde{\Psi}_{i_{k}}(t)\hat{a}_{k}e^{-i\mathbf{k.x}}+\tilde{\Psi}_{i_{k}}^{*}(t)\hat{a}_{k}^{\dagger}e^{i\mathbf{k.x}}\right] \, ,
\end{equation}
where $\hat{a}_{k}$ and $\hat{a}_{k}^{\dagger}$ are the annihilation and creation quantum operators, respectively. In addition, we rescale the Fourier mode of the spectator field as $\bar{\Psi}_{i_{k}}\equiv a^{3/2}\tilde{\Psi}_{i_{k}}$, and consequently arrive at the following evolution equation \cite{Ashoorioon:2019kcy}
\begin{equation}
\label{ddotPsibik}
\ddot{\bar{\Psi}}_{i_{k}}+\omega_{k}^{2}\bar{\Psi}_{i_{k}}=0 \, .
\end{equation}
The time-dependent angular frequency in this equation is given by
\begin{equation}
\label{omegak}
\omega_{k}^{2}\equiv\frac{k^{2}}{a^{2}}+\frac{M_{\Psi}^{2}}{\Omega^{2}}-\frac{3}{4}H^{2}-\frac{3}{2}\frac{\ddot{a}}{a}+3H\frac{\dot{\Omega}}{\Omega}-2\left(\frac{\dot{\Omega}}{\Omega}\right)^{2}+\frac{\ddot{\Omega}}{\Omega} \, .
\end{equation}
To solve the second-order different equation \eqref{ddotPsibik}, we impose the Bunch-Davies vacuum initial condition on the rescaled spectator mode at the start of preheating,
\begin{equation}
\label{PsibikBD}
\bar{\Psi}_{i_{k}}\to\frac{1}{\sqrt{2\omega_{k}}}e^{-i\omega_{k}t} \, .
\end{equation}
The solution of the differential equation \eqref{ddotPsibik} is usually used to calculate the number density of the produced particles which is given by
\begin{equation}
\label{nk}
n_{k}=\frac{\omega_{k}}{2}\left(\frac{1}{\omega_{k}^{2}}\left|\dot{\bar{\Psi}}_{i_{k}}\right|^{2}+\left|\bar{\Psi}_{i_{k}}\right|^{2}\right)-\frac{1}{2} \, .
\end{equation}
To solve the differential equation \eqref{ddotPsibik}, the value of the comoving wavenumber $k$ should be determined, and in our work, we take it as $k = 0$. The number density is an important criterion that implies the extent of the efficiency of the preheating process.

Due to quantum fluctuations of the scalar fields during preheating the effective potential acquires the following correction \cite{Kofman:1995fi, Felder:2000sf}
\begin{equation}
\label{DeltaU}
\Delta U=\frac{1}{2}M_{\tilde{\Psi}}^{2}\left\langle \tilde{\Psi}_{i}^{2}\right\rangle  \, .
\end{equation}
In our work, to calculate the dispersion of the spectator field, $\left\langle \tilde{\Psi}_{i}^{2}\right\rangle$, we perform a lattice simulation by using the computational code LATTICEEASY \cite{Felder:2000hq}. Including the above correction term, the effective potential will take the following form
\begin{equation}
\label{Utot}
U_{\mathrm{tot}}(\chi)=U(\chi)+\Delta U \, .
\end{equation}
The correction term in the potential energy may have a considerable impact on the dynamics of the Universe during preheating. This term may even cause the Universe undergoes a transient accelerating expansion during preheating, and this in turn may have cosmological outcomes. The effect of this correction on preheating will be examined in detail in the next section for a special scenario of the non-$\mathbb{M}$-flation model.


\section{Transient acceleration during preheating}
\label{section:transient_acceleration}

In the previous section, we have presented the basic equations required for the investigation of preheating in the framework of non-$\mathbb{M}$-flation. In this section, we consider a specific case of this setting and use those equations to investigate the preheating process for it.

It should be noted that taking non-vanishing values for the parameter $\kappa_{\mathrm{eff}}$ leads to some problems. If we take this parameter as $\lambda m^{2}=4\kappa^{2}/9$ which restricts the inflaton dynamics to the super-gravity equations, then the model fails to provide a successful process of preheating in the case that the inflaton field never reaches the super-symmetry minimum. This case will be discussed in detail in Appendix \ref{appendix:symmetry-breaking}. In addition, in the cases for which the inflaton can cross the potential bump and reach the super-symmetry minimum, then the effective mass squared of the preheat field will be negative in a range of $\chi$. This in turn cause that after inclusion of the potential correction term in the background dynamics, the effective potential of inflaton acquires negative values in that range of $\chi$. The negative values of the effective potential lead to some critical problems in numerical computations, and therefore here we focus on the case $\kappa_{\mathrm{eff}}=0$ to prevent the numerical errors in our computations.

Also, it should be noted that if we take $\mu\equiv\sqrt{2/\lambda_{\mathrm{eff}}}\,m\gtrsim0.001M_{P}$, then the dispersion of the Einstein-frame preheat field, $\left\langle \tilde{\Psi}_{i}^{2}\right\rangle $, gets very large values which are even of order $M_P$ or more. This means that the magnitude of quantum fluctuations of the preheat field becomes of the order of the inflaton field magnitude, and this, in turn, makes some problems in the LATTICEEASY code. Therefore, in order to do a successful lattice simulation, we have taken $\mu=10^{-4}M_{P}$ which is small enough to avoid such problems. For the coupling constant to gravity, we choose $\xi=100$ which is large enough to make the inflationary observables compatible with the current CMB observations. We further take $\omega=-20$ which is related to $\alpha$-mode with $j = 18$.

Since we have set $\kappa_{\mathrm{eff}}=0$ here, then the potential in both the Jordan and Einstein frames has only one minimum. In the Jordan frame, the potential  is obtained from Eq. \eqref{V0} as
\begin{equation}
\label{V0-2}
V_{0}(\phi)=\frac{\text{\ensuremath{\lambda}}_{\mathrm{eff}}}{4}\phi^{2}\left(\phi^{2}+\mu^{2}\right) \, .
\end{equation}
Substituting this into Eq. \eqref{U}, the Einstein-frame potential is obtained as
\begin{equation}
\label{U-2}
U(\chi)=\frac{\text{\ensuremath{\text{\ensuremath{\lambda}}_{\mathrm{eff}}}}\phi^{2}(\chi)\left[\phi^{2}(\chi)+\mu^{2}\right]}{4\left[\xi\phi^{2}(\chi)+1\right]^{2}} \, .
\end{equation}
The potential of this case is quite symmetric around its minimum in both the Jordan and Einstein frames. Because of the symmetry of this potential, both the left and right branches of the potential give rise to similar inflationary scenarios. However,  here we concentrate on the inflationary trajectory from the right branch of the potential. From fixing the amplitude of the scalar perturbations according to the recent CMB observations, we find $\text{\ensuremath{\lambda}}_{\mathrm{eff}}=3.780\times10^{-6}$. This value of $\text{\ensuremath{\lambda}}_{\mathrm{eff}}$ gives the number of simultaneous branes as $N\approx130$ which is much less than what would be required in the M-flation scenario. We further get the scalar spectral index and tensor-to-scalar ratio, respectively, as $n_{s}=0.9678$ and $r=0.003$ which are in good agreement with the Planck 2018 observations \cite{Planck:2018jri}.

To do the numerical commutations for this case, it is convenient to use the approximated function in Eq. \eqref{phi-chi-approx} to express $\phi$ in terms of $\chi$. The constant coefficients of this function for this case are presented in Table \ref{table:a-b}. Also, it is much appropriate to apply approximation \eqref{MPsit2-approx} for $M_{\tilde{\Psi}}^{2}$ in the code. In the following, we will verify explicitly that this approximation is really valid in our examination. In our lattice simulation, we have set the number of grid points per edge of the cubical lattice as $N = 128$. The total number of points on the lattice therefore will be $128^3$. We also set the size of the box (i.e. length of each edge) in rescaled distance units as $L = 40$, and so the volume of the box will be $40^3$. The size of the box in the program units, $L$, is related to the physical size of the box, $L_{\mathrm{phys}}$, via $L=\tilde{m}L_{\mathrm{phys}}$, where $\tilde{m}$ is a mass parameter which is usually taken to be of the order of the inflaton effective mass, and it has been set to $10^{-6}m_{P}$ in our computations.

\begin{table}[t]
\centering
\scalebox{1.0}{
\begin{tabular}{|c|c|c|c|}
\hline
$a_{1}$ & $9.263\times10^{-1}$ & $b_{1}$ & $-3.343\times10^{-7}$\tabularnewline
\hline
$a_{2}$ & $-2.562\times10^{-7}$ & $b_{2}$ & $8.087\times10^{3}$\tabularnewline
\hline
$a_{3}$ & $4.042\times10^{3}$ & $b_{3}$ & $-5.910\times10^{-4}$\tabularnewline
\hline
$a_{4}$ & $-1.802\times10^{-4}$ & $b_{4}$ & $3.539\times10^{6}$\tabularnewline
\hline
$a_{5}$ & $9.761\times10^{5}$ & $b_{5}$ & $1.367\times10^{-2}$\tabularnewline
\hline
$a_{6}$ & $9.271\times10^{-3}$ & $b_{6}$ & $1.286\times10^{8}$\tabularnewline
\hline
$a_{7}$ & $2.079\times10^{7}$ & $b_{7}$ & $2.072$\tabularnewline
\hline
$a_{8}$ & $3.344\times10^{-1}$ & $b_{8}$ & $4.225\times10^{8}$\tabularnewline
\hline
$a_{9}$ & $4.102\times10^{7}$ & $b_{9}$ & $5.963$\tabularnewline
\hline
$a_{10}$ & $5.548\times10^{-1}$ & $b_{10}$ & $8.054\times10^{6}$\tabularnewline
\hline
\end{tabular}
}
\caption{The constant coefficients in function \eqref{phi-chi-approx} which is used in our investigation to approximate the Jordan-frame scalar field $\phi$ in terms of the one in the Einstein frame $\chi$.}
\label{table:a-b}
\end{table}

\begin{figure}
\centering
\includegraphics[scale=0.4]{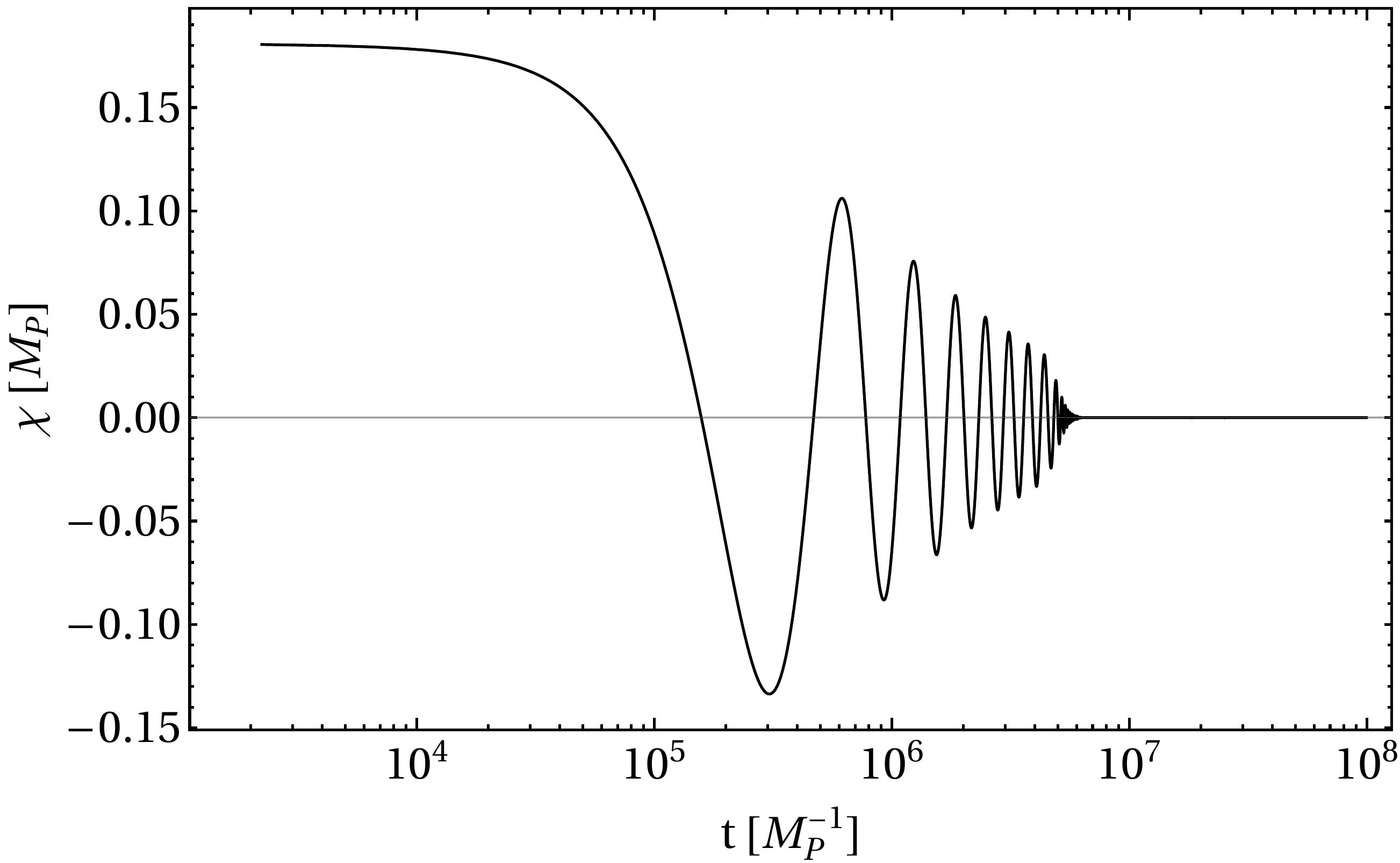}
\caption{The variation of the Einstein-frame inflaton field ($\chi$) versus time in our setup.}
\label{figure:chi-t-2}
\end{figure}

\begin{figure}
\centering
\includegraphics[scale=0.4]{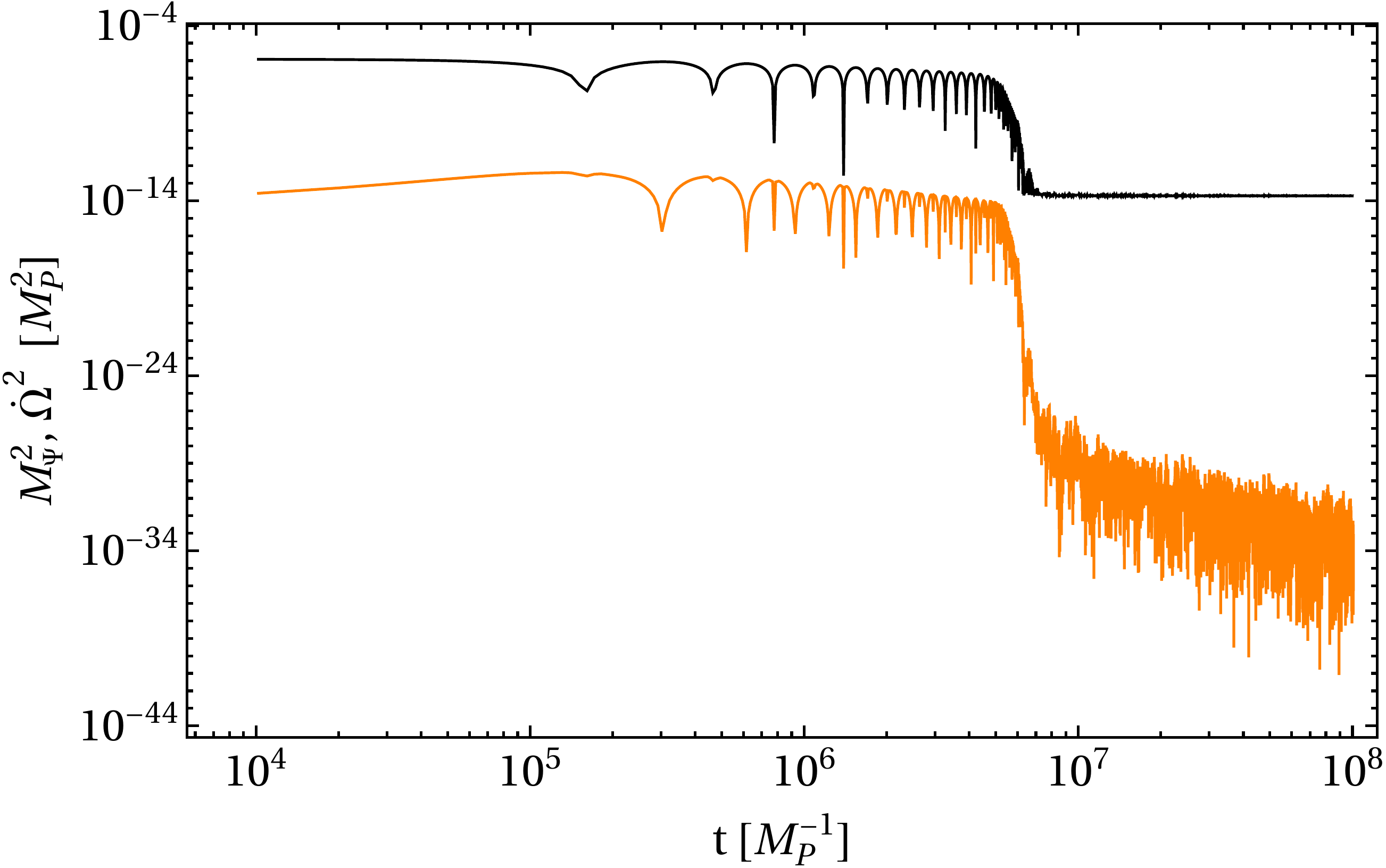}
\caption{The evolution of $M_{\Psi}^{2}$ (black) and $\dot{\Omega}^{2}$ (orange) with time in our model.}
\label{figure:MPsi2_Omegadot2-2}
\end{figure}

Using LATTICEEASY, the evolution of the Einstein-frame scalar field $\chi$ with time is obtained as shown in Fig. \ref{figure:chi-t-2}. We see in the figure that the $\chi$ field oscillates around the potential minimum which is located at $\chi = 0$, and as time evolves, the amplitude of the oscillations reduces. Having the time evolution of $\chi$ in hand, we are able to draw the diagram of the evolution of $M_{\Psi}^{2}$ and $\dot{\Omega}^{2}$ vs. time as illustrated in Fig. \ref{figure:MPsi2_Omegadot2-2}. It is evident from this figure that during the examined time interval, the inequality $\dot{\Omega}^{2}\ll M_{\Psi}^{2}$ is indeed valid, and therefore approximation \eqref{MPsit2-approx} that we used for $M_{\tilde{\Psi}}^{2}$ in our calculations, is completely viable.

\begin{figure}
\centering
\includegraphics[scale=0.4]{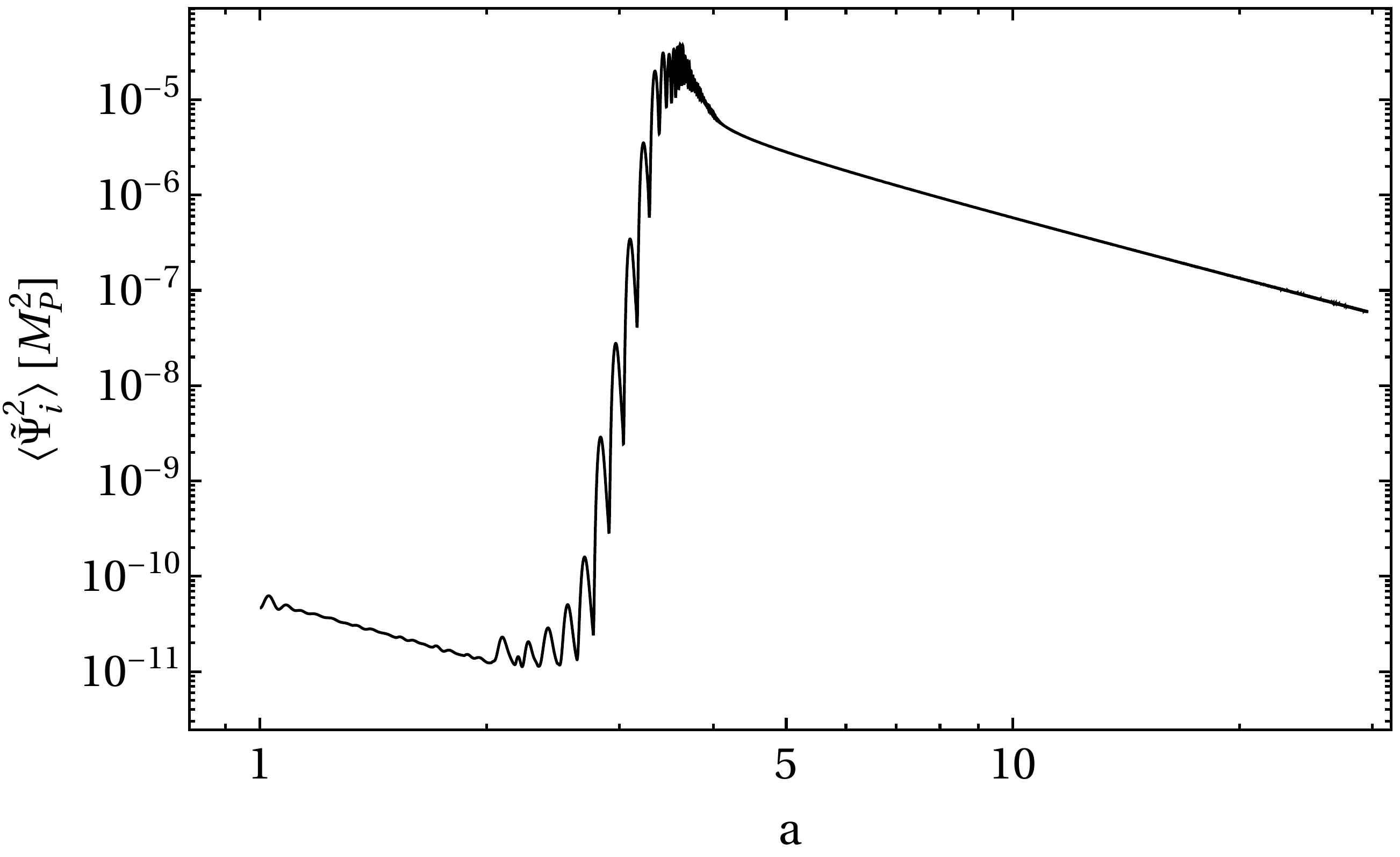}
\caption{The result of the LATTICEEASY code for the variation of the dispersion of the Einstein-frame preheat field, $\left\langle \tilde{\Psi}_{i}^{2}\right\rangle $, in terms of scale factor in our framework.}
\label{figure:avPsiti2-2}
\end{figure}

The result of the LATTICEEASY code for the dispersion of the Einstein-frame preheat field, $\left\langle \tilde{\Psi}_{i}^2\right\rangle $, in terms of the scale factor, is presented in Fig. \ref{figure:avPsiti2-2}. Throughout this work, we normalize the scale factor to its value at the start of our simulation which is very close to the scale factor of the Universe at the end of inflation. The figure shows that after a period of time, $\left\langle \tilde{\Psi}_{i}\right\rangle $ begins to grow an then reach its maximum value $\left\langle \tilde{\Psi}_{i}^{2}\right\rangle \approx4\times10^{-5}$ at the scale factor $a\approx3.6$. After that, it decreases with scale factor, and for $a\gtrsim5$, it declines monotonically like $\left\langle \tilde{\Psi}_{i}^{2}\right\rangle \propto a^{-2}$.

\begin{figure}
\centering
\includegraphics[scale=0.4]{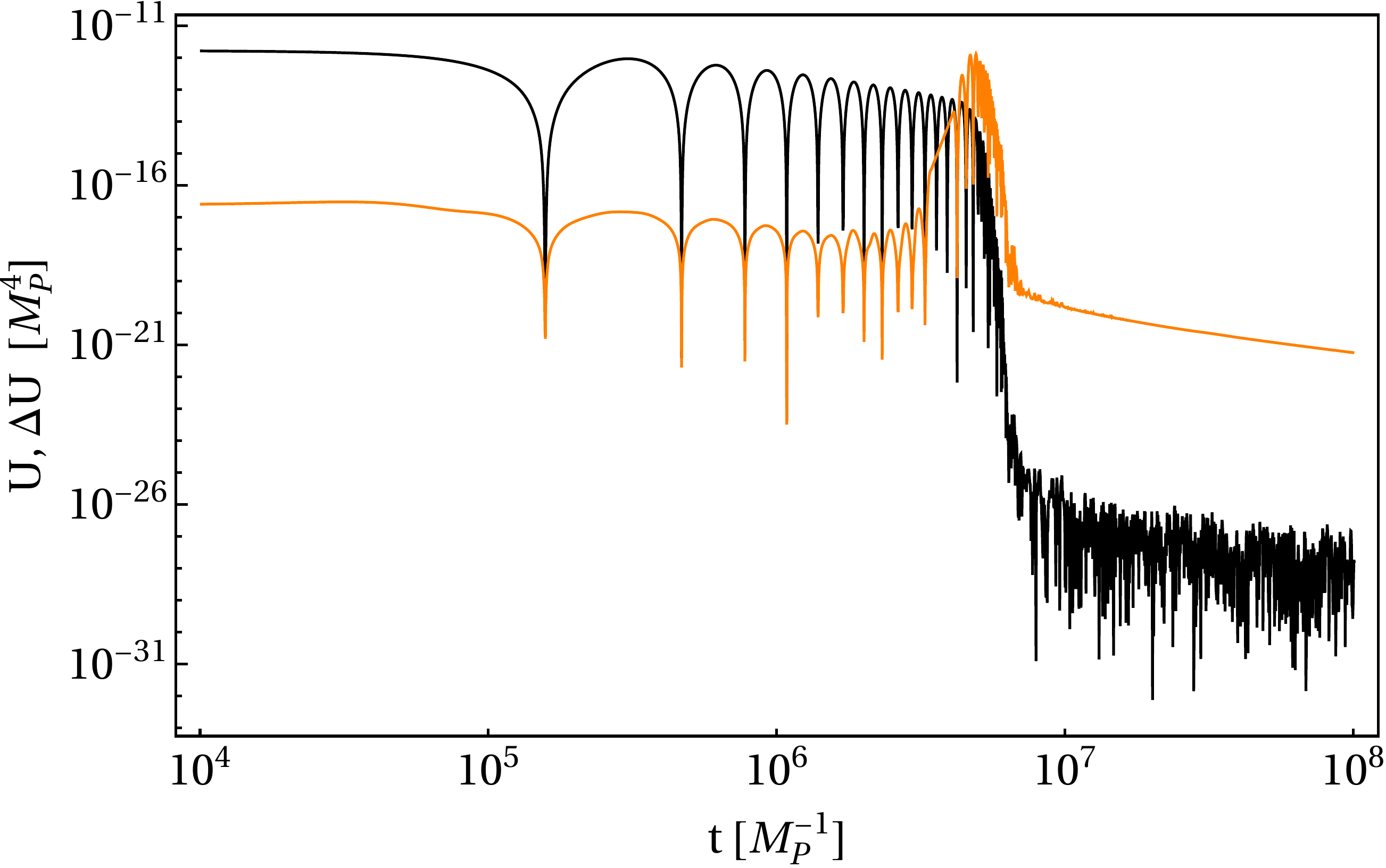}
\caption{The evolution of the potential correction term $\Delta U$ (orange) and the original potential $U$ (black) versus time in the investigated model.}
\label{figure:U_deltaU-t-2}
\end{figure}

\begin{figure}
\centering
\includegraphics[scale=0.34]{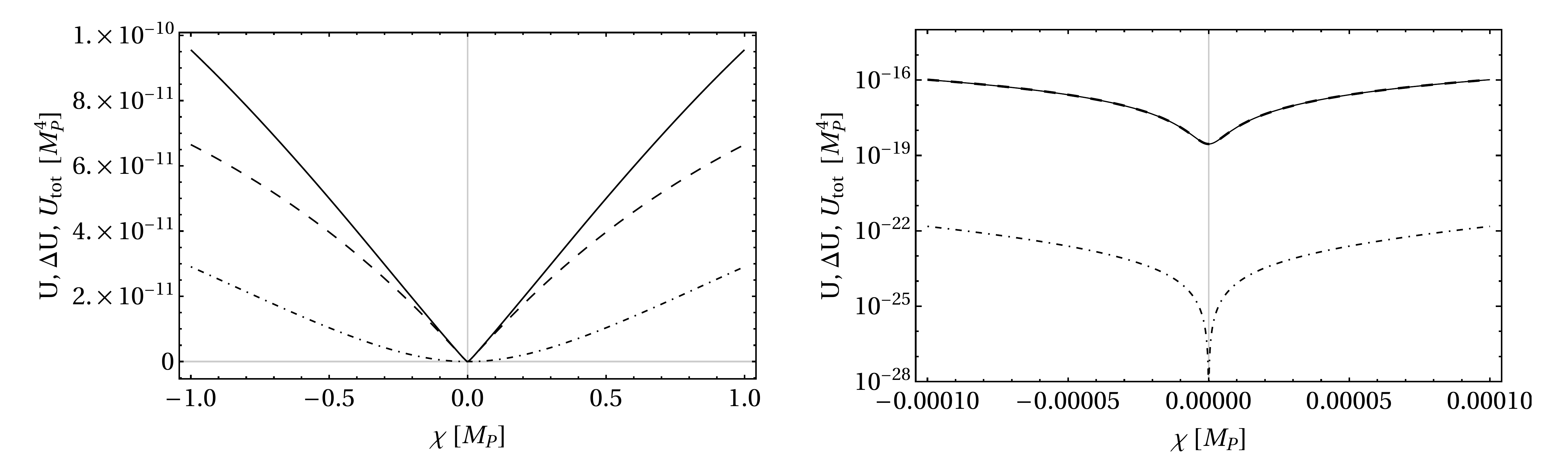}
\caption{The modified potential $U_{\mathrm{tot}}(\chi)$ (solid line) in comparison with the original potential $U(\chi)$ (dash-dotted line). The correction term $\Delta U$ also has been shown in the figure by the dashed line. The plot in the left panel has been drawn in the usual linear scale, while the one in the right panel has been drawn in the logarithmic scale for a better demonstration.}
\label{figure:U-2}
\end{figure}

By using the results for the time variation of $\left\langle \tilde{\Psi}_{i}^{2}\right\rangle$ in Eq. \eqref{DeltaU}, we can calculate the potential correction term $\Delta U$ and include it in the background dynamics at each instant of time. Using the result for $\Delta U$ in Eq. \eqref{Utot}, we can evaluate the modified potential $U_{\mathrm{tot}}$. The time variation of $\Delta U$ in comparison with the one of the original potential $U$ has been depicted in Fig. \ref{figure:U_deltaU-t-2}. The figure implies that around the time $t\approx4.3\times10^{6}M_{P}^{-1}$, the correction term $\Delta U$ dominates over the original potential $U$, and its domination continues in all of the subsequent times. Due to this domination, the Universe will undergo a transient accelerated expansion during preheating, before it settles into a decelerating phase again. The diagram of $\Delta U$ represents a peak with the value $\Delta U=1.277\times10^{-5}M_{P}^{4}$ at the time $t=4.915\times10^{6}M_{P}^{-1}$. At this moment, the vev of the preheat field squared gets the value of $\left\langle \tilde{\Psi}_{i}^{2}\right\rangle =3.021\times10^{-5}M_{P}^{2}$. By using this value in Eqs. \eqref{DeltaU} and \eqref{Utot}, we can draw the diagram of the potential correction term $\Delta U$ and the modified potential $U_{\mathrm{tot}}(\chi)$ at that moment as presented in Fig. \ref{figure:U-2}. In the figure, we have also compared the modified potential with the original potential $U(\chi)$. The effect of the correction term can be seen more clearly in the right panel of the figure. The figure shows obviously that the correction term $\Delta U$ lifts the potential so that it takes a positive value at its minimum while the original potential vanishes at that point.

\begin{figure}
\centering
\includegraphics[scale=0.4]{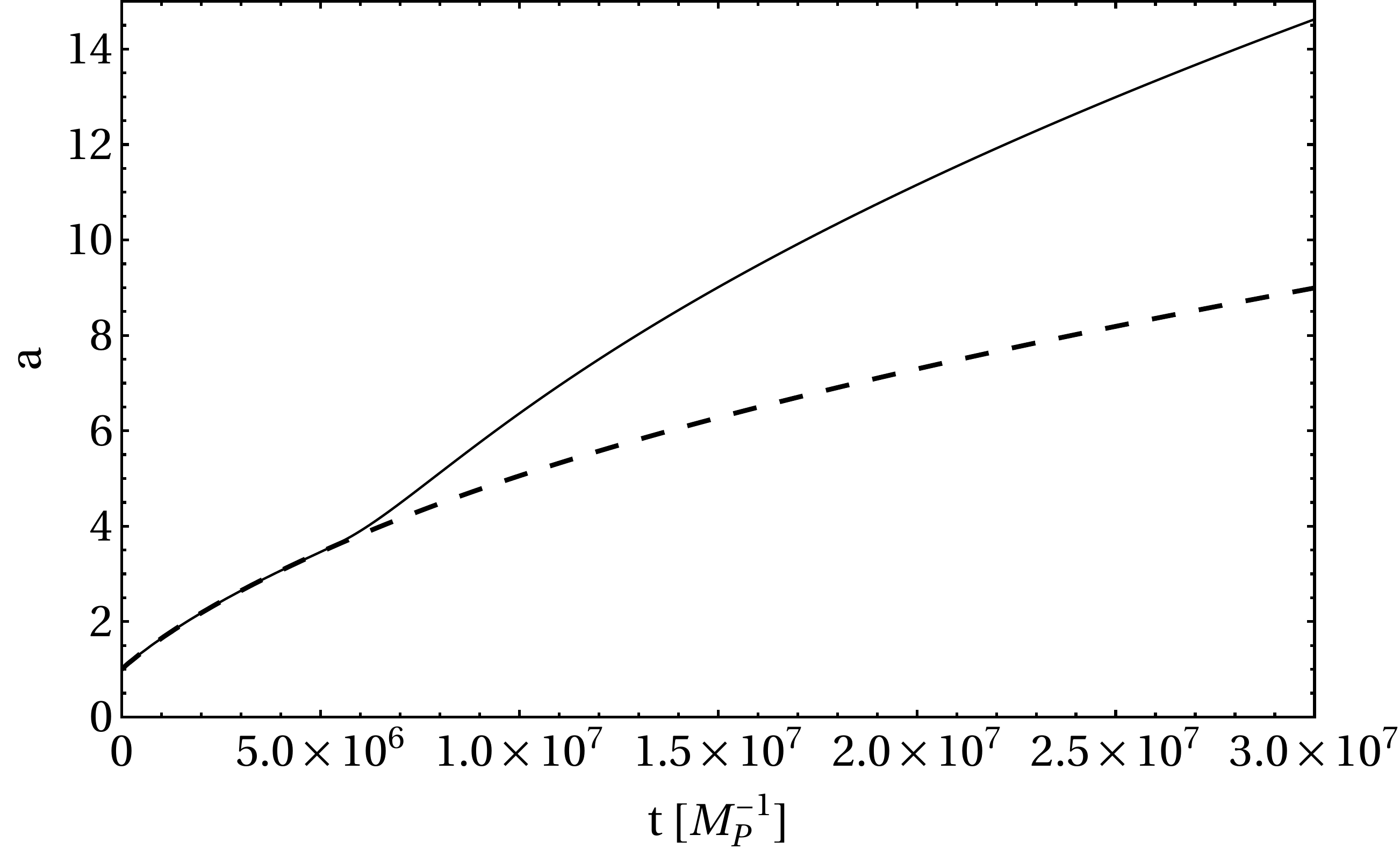}
\caption{The evolution of scale factor with time in our scenario. The solid line shows the result by including $\Delta U$, while the dashed line shows the one without including it.}
\label{figure:a-2}
\end{figure}

\begin{figure}
\centering
\includegraphics[scale=0.4]{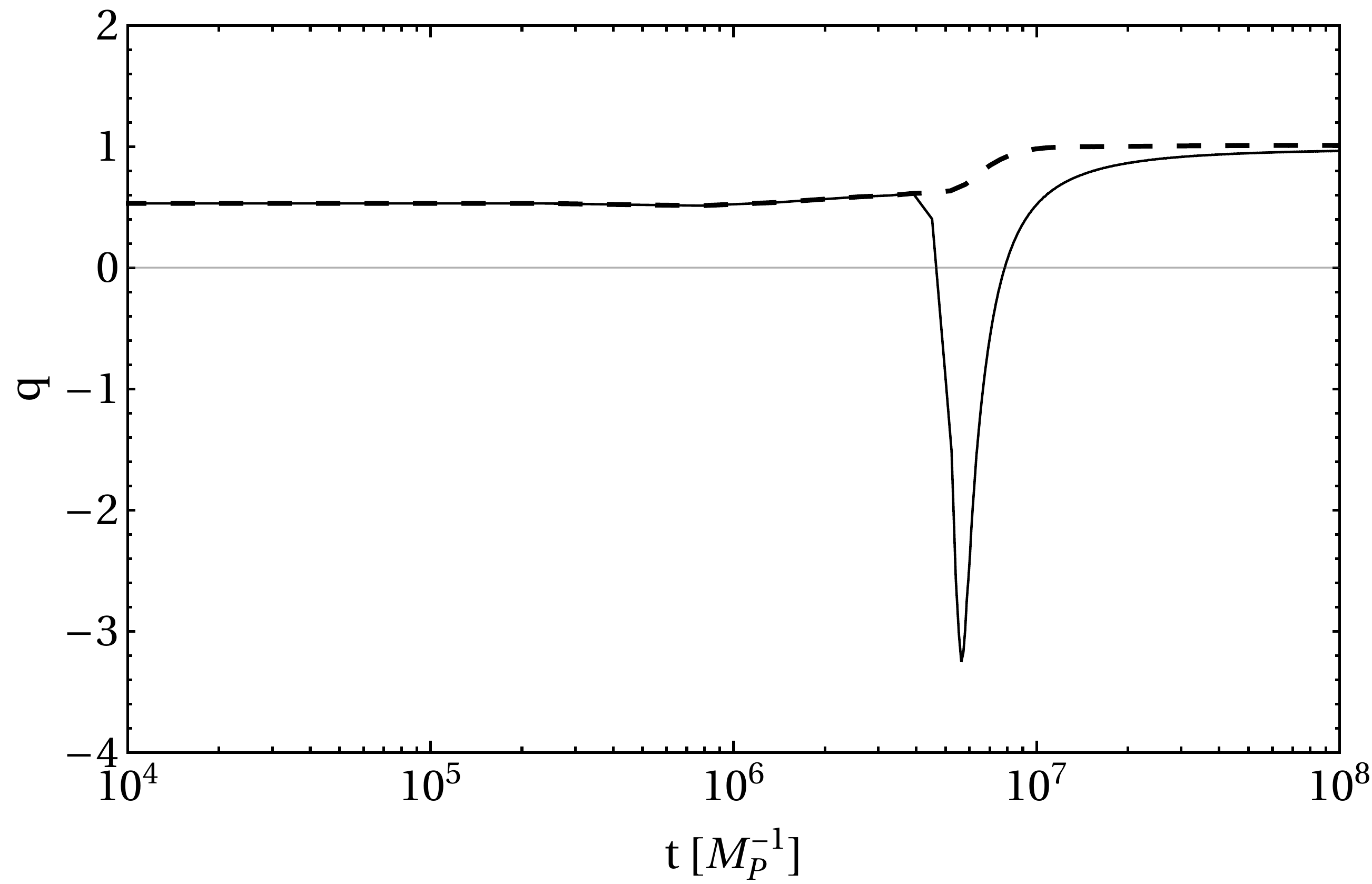}
\caption{The deceleration parameter $q\equiv-a\ddot{a}/\dot{a}^{2}$ with respect to cosmic time in our model. The solid line corresponds to the case with including $\Delta U$, while the dashed line is corresponding to the case without including it.}
\label{figure:q-2}
\end{figure}

During a period of time around the instance $t=4.915\times10^{6}M_{P}^{-1}$ that the correction term takes rather considerable values, the effect of the lift in the inflaton potential becomes decisive in the background dynamics of the Universe. In this period, the non-vanishing value at the potential minimum causes the Universe to exit temporarily from the deceleration phase and enter an acceleration phase of expansion. The change of behavior from the deceleration to acceleration can be seen in Fig. \ref{figure:a-2} which demonstrates the variation of the scale factor with time in this case. The transition from deceleration to a temporary acceleration can also be seen more clearly in Fig. \ref{figure:q-2} that displays the plot of the deceleration parameter $q\equiv-a\ddot{a}/\dot{a}^{2}$ versus time. To avoid the oscillations of this parameter, here we have calculated the average of this parameter on each cycle of the $\chi$ field oscillation. The solid curve in the figure corresponds to the case with the inclusion of $\Delta U$, while the dashed curve is related to the case without the inclusion of the correction term. In the case that $\Delta U$ is not included in the evolution of the background, the deceleration parameter is always positive which indicates the expansion of the Universe remains always decelerating. However, in the case that $\Delta U$ was included, the parameter $q$ can get negative values in an interval of time. Before the time $4.6\times10^{6}M_{P}^{-1}$, we have $q\approx0.5>0$ and the Universe is in a phase of deceleration. Between the times $4.6\times10^{6}M_{P}^{-1}$ and $7.8\times10^{6}M_{P}^{-1}$, we have $q<0$ which indicates that the Universe experiences an accelerating phase. After $t\approx7.8\times10^{6}M_{P}^{-1}$, the Universe transits again to a deceleration phase with $q>0$. The occurrence of a phase of temporary acceleration after inflation may have notable consequences on the theoretical and observational implications. For example, this stage of nonthermal acceleration during preheating may be useful in the dilution of moduli and gravitino which otherwise lead to problems \cite{Felder:2000sf}.

\begin{figure}
\centering
\includegraphics[scale=0.4]{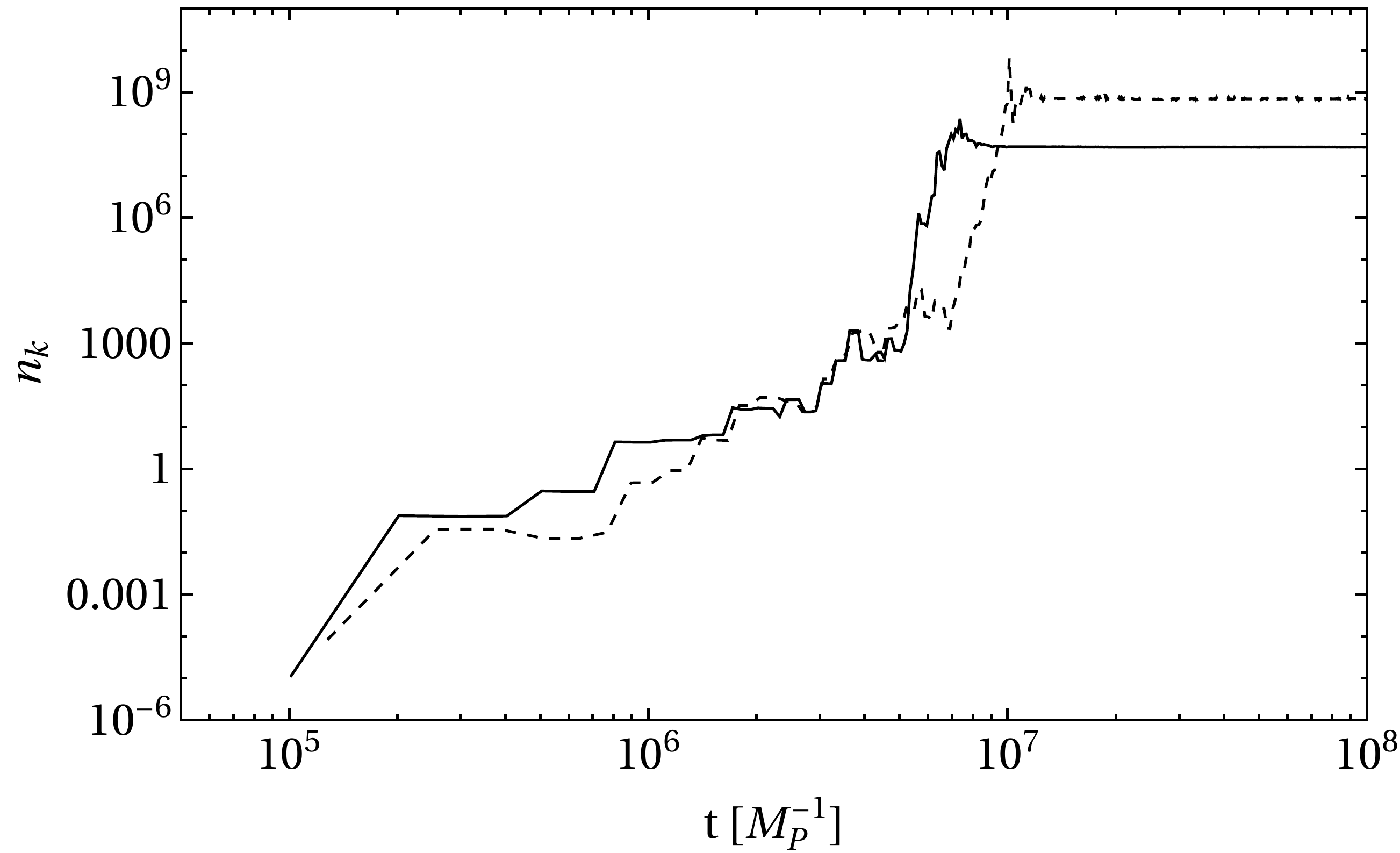}
\caption{The variation of number density with time in our investigation. The dashed plot shows the number density without the inclusion of the corrected term $\Delta U$, while its contribution has already been involved in the solid plot.}
\label{figure:nk-2}
\end{figure}

In Fig. \ref{figure:nk-2}, we have examined the impact of the correction term to the inflaton potential on the number density \eqref{nk} of the particles generated during preheating. In this figure, the dashed and solid plots correspond to the case with the inclusion of $\Delta U$ and the case without the inclusion of it, respectively. If we compare these plots with each other, we see that in the presence of $\Delta U$, the number density takes smaller values, but the efficiency of preheating is still enough to explain the process of particle production after inflation. Therefore, we conclude that consideration of the potential correction term decreases the efficiency of the preheating mechanism to some extent.

\section{Gravitational waves from preheating}
\label{section:GWs}

To estimate the spectrum of the stochastic GW signal from preheating, we apply a modified version of LATTICEEASY based on the equations of \cite{Dufaux:2007pt} to evaluate tensor perturbations in the post-inflationary era\footnote{We thank G. Felder for sending us this code version}. We have verified that this version works properly and is able to reproduce previous results. In particular, we have reproduced the results of \cite{Easther:2006vd} with this version of the code. We have implemented our non-minimal M-flation model in this code, and our modified version is available online\footnote{\url{https://github.com/krezazadeh/latticeeasy2.1-non-minimal_M-flation}}. In the approach of this code, a Green function method is used to evaluate the metric inhomogeneities on a lattice. In this formulation, the present-day frequency and amplitude of the GWs spectrum are given respectively by \cite{Dufaux:2007pt}
\begin{align}
f &= \frac{k}{a_{j}\rho_{j}^{1/4}}\left(\frac{a_{j}}{a_{*}}\right)^{1-\frac{3}{4}(1+w)}4\times10^{10}\,\mathrm{Hz} \, ,
\label{f}
\\
\Omega_{GW}h^{2} &= \frac{S_{k}\left(\tau_{f}\right)}{a_{j}^{2}\rho_{j}}\left(\frac{a_{j}}{a_{*}}\right)^{1-3w}\left(\frac{g_{*}}{g_{0}}\right)^{-1/3}\Omega_{r}h^{2} \, .
\label{OmegaGWh2}
\end{align}
Here, $\tau_f$ denotes the conformal time at the end of the simulation that is the moment at which the GWs are generated during preheating. In addition, $a_j$ denotes the scale factor at the moment that the equation of state parameter $w$ jumps to $w = 1/3$, and $a_*$ represents the scale factor at the moment when thermal equilibrium is established. It should be noted that $a_j$ and $a_*$ cannot be determined in our investigation, because they depend on the subsequent interactions of the fields which are included in our analysis. To overcome this problem, we follow the steps of \cite{Easther:2006vd}, and assume that the Universe enters the radiation-dominated era after the end of the simulation. In this way, the values of $a_j$ and $a_*$ are assumed to be very close to the scale factor at the end of the simulation. For the abundance of radiation today we take $\Omega_{{}_{\rm r}}h^{2}\approx4.3\times10^{-5}$. Furthermore, for the ratio of the number of degrees of freedom at matter-radiation equality to the number of degrees of freedom today, we take $g_*/g_0=1/100$. The function $S_{k}\left(\tau_{f}\right)$ in \eqref{OmegaGWh2} has relation with transverse-traceless part of the energy-momentum $T_{ij}^{TT}$ as
\begin{align}
S_{k}\left(\tau_{f}\right)= & \frac{4\pi Gk^{3}}{V}\int d\Omega\sum\limits _{i,j}\Bigg\{\left|\int_{\tau_{i}}^{\tau_{f}}d\tau'\cos\left(k\tau'\right)a\left(\tau'\right)T_{ij}^{TT}\left(\tau',\mathbf{k}\right)\right|^{2}
\nonumber
\\
& +\left|\int_{\tau_{i}}^{\tau_{f}}d\tau'\sin\left(k\tau'\right)a\left(\tau'\right)T_{ij}^{TT}\left(\tau',\mathbf{k}\right)\right|^{2}\Bigg\} \, .
\label{Sk}
\end{align}
This is the main quantity that is computed numerically on the lattice.

\begin{figure}
\centering
\includegraphics[scale=0.4]{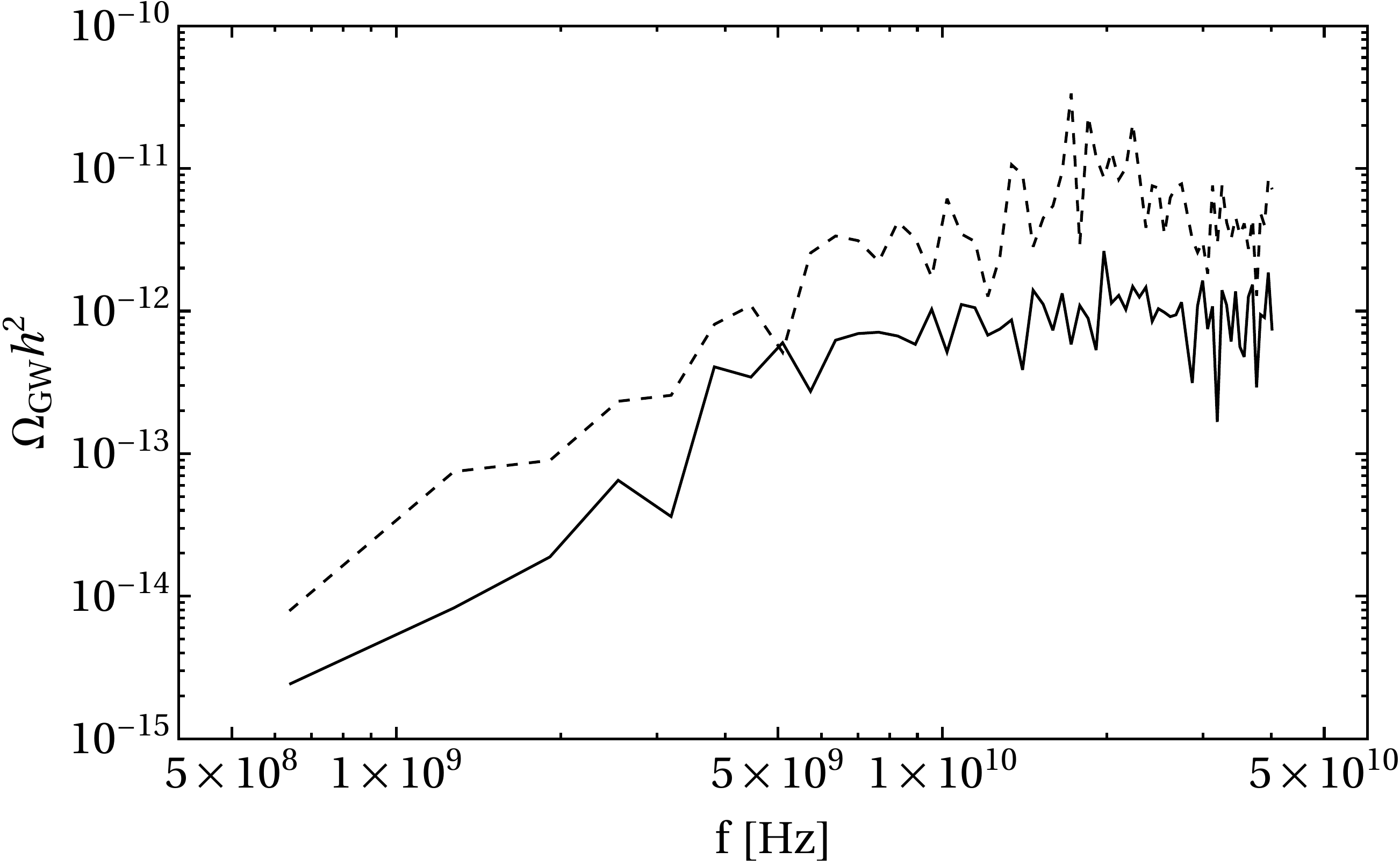}
\caption{The LATTICEEASY results for the spectrum of the GWs produced during preheating in our setup. The solid plot shows the spectrum of the case with the inclusion of the potential term $\Delta U$, while the dashed plot corresponds to the case without involving it in the background dynamics. The spectrum in both cases has been evaluated at the epoch when $H=5.5\times10^{-9}M_{P}$.}
\label{figure:OmegaGWh2-2}
\end{figure}

The spectrum of GWs provided by LATTICEEASY for this model has been demonstrated in Fig. \ref{figure:OmegaGWh2-2}. In the figure, the spectrum is evaluated at the moment when the Hubble parameter is equal to $H=5.5\times10^{-9}M_{P}$. We have followed this convention because the amplitude of the tensor perturbations is strongly connected to the energy scale of the Universe, and it in turn is directly related to the Hubble parameter through the Friedmann equation. The results of our computations indicate that after a time in our simulation, the spectra of the GWs accumulate on each other, and the frequencies and amplitudes remain almost constant. This ensures that the code has converged appropriately enough in our simulation. The solid curve in this figure shows the spectrum for the case where $\Delta U$ is included, whereas the dashed one shows the spectrum without it. In the case where $\Delta U$ is included, the peak of the spectrum is of order $\Omega_{\mathrm{GW}}h^{2}\sim2.6\times10^{-12}$, but in the case where correction term was not included, the peak is of order $\Omega_{\mathrm{GW}}h^{2}\sim3.2\times10^{-11}$. So the amplitude of GWs produced during preheating is reduced by an order of magnitude. For both cases, the spectrum appears in the frequency range $f\sim6\times10^{8}-4\times10^{10}\,\mathrm{Hz}$. The range of the frequency and amplitude of the spectrum in the case without the correction term is almost similar to those values in the benchmark setup of preheating \cite{Easther:2006vd}, but the amplitude of the spectrum in the case with the involving it is one order of magnitude less than the result of the benchmark scenario. We see that the inclusion of the correction term reduces the amplitude of the stochastic GWs generated during preheating by one order of magnitude compared to the case without it. The GW signal could be possibly detected by futuristic experiments through the conversion of propagating GW through an electromagnetic field that sources a feeble electromagnetic which oscillates with the frequency of the propagating electromagnetic field. If the GWs are coherent on many wave cycles, resonant detectors can detect such a signal. In fact detectors like ADMX \cite{ADMX:2021nhd}, HAYSTAC\cite{HAYSTAC:2018rwy}, ORGAN\cite{McAllister:2017lkb}, and CAPP\cite{CAPP:2020utb} uses such a mechanism to detect ultralight axions. The resonant frequency of such detectors with a geometric size $L_{\rm det}\sim \mathcal{O}(\mathrm{cm})-\mathcal{O}(\mathrm{m})$ makes them sensitive to GWs in the GHz regime.


\section{Conclusions}
\label{section:conclusions}

Light preheat fields coupled to the inflaton, can drain the energy of the inflaton during the deceleration phase. However as the inflaton energy is pumped to the preheat field and it obtains nonzero number densities, nonzero dispersion for the preheat field is developed that can affect the preheating phase. Here we investigated this issue in the context of non-symmetry-breaking non-minimal inflation. The M-flation scenario has motivations from the string theory, and it can resolve some of the problems of the chaotic inflationary models. However, despite its successes, M-flation suffers from several shortcomings that challenge its foundation. One way to remedy the problems of M-flation is the setup of non-$\mathbb{M}$-flation in which the inflaton field couples non-minimally to gravity. In this framework, we can improve the consistency of the inflationary observables with the recent CMB observations, in comparison with the minimal M-flation. The number of the D3-branes in this framework could be considerably less than the required number in the minimal setup, which ameliorates the issue of backreaction.

The non-$\mathbb{M}$-flation model, like the minimal M-flation scenario, has this interesting feature: it possesses scalar and gauge spectator fields. These fields are frozen classically during inflation, but they can play an important role in the dynamics of the Universe in the preheating era after inflation. The quantum vev of the spectator field squared leads to additional energy which lifts the potential energy of the inflaton field. We examined the effects of the modified potential on the dynamics of the post-inflationary Universe in the setting of non-$\mathbb{M}$-flation. In this framework, we focused on the non-symmetry-breaking model in which the potential has only one minimum in its shape in both the Jordan and Einstein frames, and the potential is completely symmetric around this minimum in both frames. This model satisfies the current CMB constraints from the Planck 2018 measurements on the inflationary observables. In addition, since the number density can grow substantially in this model, therefore it provides an appealing preheating mechanism.

We used the LATTICEEASY code to compute the dispersion of the Einstein-frame preheat field, $\left\langle \tilde{\Psi}_{i}^2\right\rangle $, at each instant of time. The results of this quantity were used to evaluate the potential correction term $\Delta U$ and also the modified potential $U_{\mathrm{tot}}$. We included the contribution of $\Delta U$ in the inflaton dynamics and examined their effects on the preheating process. Our findings imply that around a special epoch during preheating, the correction term $\Delta U$ dominates over the original potential $U$, and it continues to be dominant in all of the subsequent times. Due to the correction term $\Delta U$, the inflaton potential is lifted so that it acquires a positive value at its minimum while the original potential vanishes at that point. The non-vanishing value at the potential minimum causes the Universe to transit to a temporary acceleration phase of expansion during its evolution in the preheating era. This transient period of acceleration is a remarkable consequence coming from consideration of the quantum correction term in the post-inflationary dynamics. The occurrence of this phase of acceleration after inflation may have notable consequences. For example, this period of nonthermal acceleration during preheating may help in solving the moduli and gravitino problems \cite{Felder:2000sf}. We furthermore examined the impact of the correction term to the inflaton potential on the number density of the particles generated during preheating in our setup. We showed that in the presence of $\Delta U$, the number density takes smaller values compared to the case without it, but the efficiency of preheating is still high enough to explain the process of particle production after inflation. Therefore, we conclude that consideration of the potential correction term reduces the efficiency of the preheating probes to some extent.

Finally, we examined the effect of the potential correction term on the stochastic GWs generated during preheating in our model. For this purpose, we applied LATTICEEASY to evaluate the spectrum of these GWs in our setup. Our results demonstrate that if the contribution of the potential correction term is included in the Universe dynamics, then the peak of the GWs spectrum becomes of order $\Omega_{\mathrm{GW}}h^{2}\sim2.6\times10^{-12}$, while in the case without involving it, the peak is of order $\Omega_{\mathrm{GW}}h^{2}\sim3.2\times10^{-11}$. For both cases, the spectrum appears in the frequency range $f\sim6\times10^{8}-4\times10^{10}\,\mathrm{Hz}$. The range of the frequency and amplitude of the spectrum of the case without the correction term is almost similar to those values in the benchmark setup of preheating \cite{Easther:2006vd}, but the amplitude of spectrum in the case with the involving $\Delta U$ is one order of magnitude less than the one in the benchmark scenario. Therefore, we conclude that the inclusion of the correction term reduces the amplitude of the stochastic GWs generated during preheating one order of magnitude compared to the case without it. In both cases the GW signal from preheating is in the GHz regime. As mentioned before, such a signal could be potentially detected by the resonant conducting cavity of a centimeter to meter size, in which the propagating GW produces a feeble EM field with the same frequency.



\appendix


\section{Preheating in a symmetry-breaking model of non-$\mathbb{M}$-flation}
\label{appendix:symmetry-breaking}

In this appendix, we study preheating within a symmetry-breaking model of non-$\mathbb{M}$-flation, where the inflaton only oscillates around the symmetry-breaking vacuum. In this model, we assume that the inflaton dynamics is restricted to satisfy the supergravity equations with a constant dilaton for which $\lambda m^{2}=4\kappa^{2}/9$. In \cite{Ashoorioon:2019kcy}, we concluded that in the non-minimal setup oscillations around this minimum can also lead to some particle production.  Hereby, we show that the conclusion that we had drawn was wrong.

\begin{figure}
\centering
\includegraphics[scale=0.4]{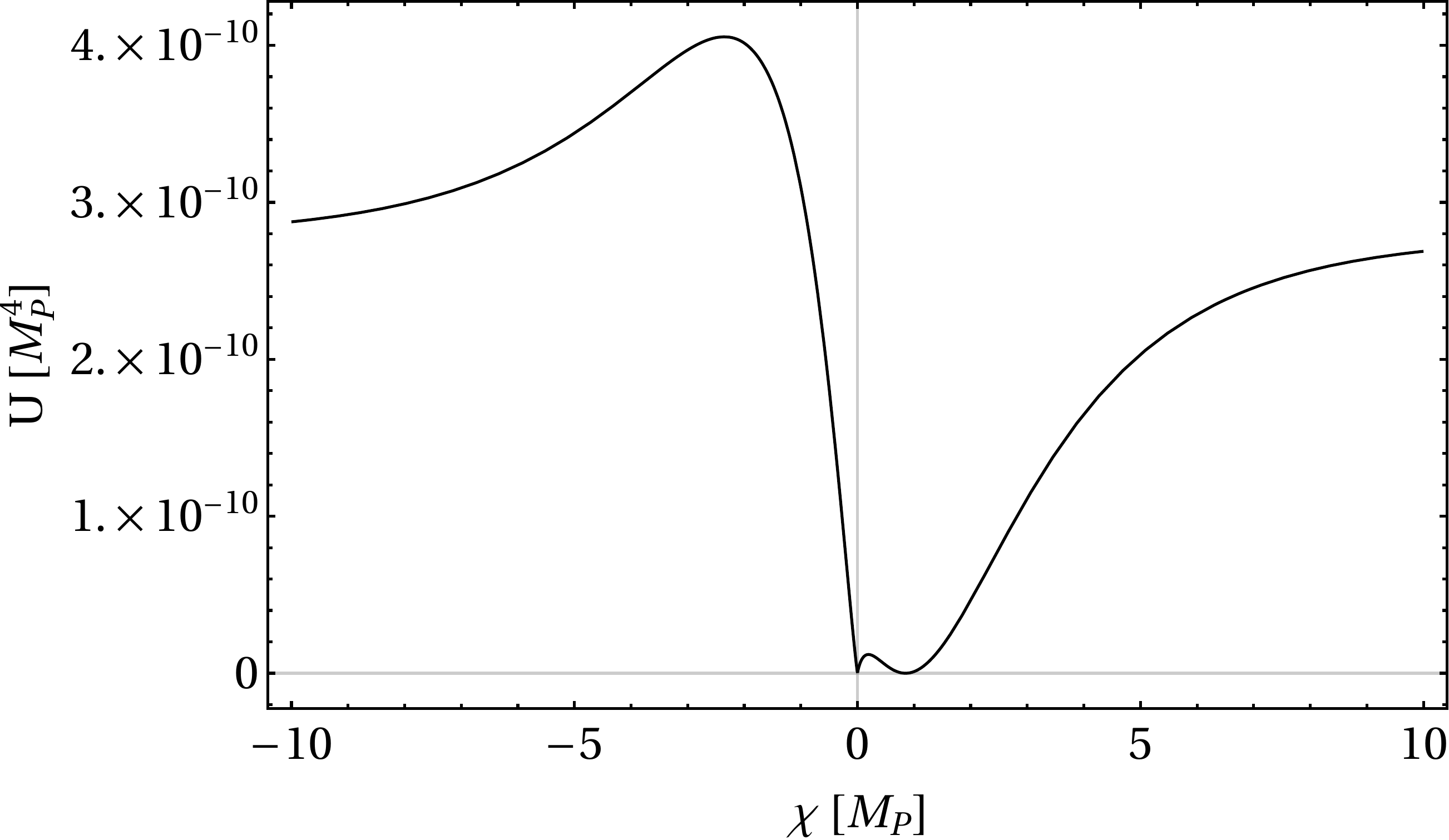}
\caption{The inflaton potential in the Einstein frame for the symmetry-breaking model.}
\label{figure:U-1}
\end{figure}

\begin{figure}
\centering
\includegraphics[scale=0.4]{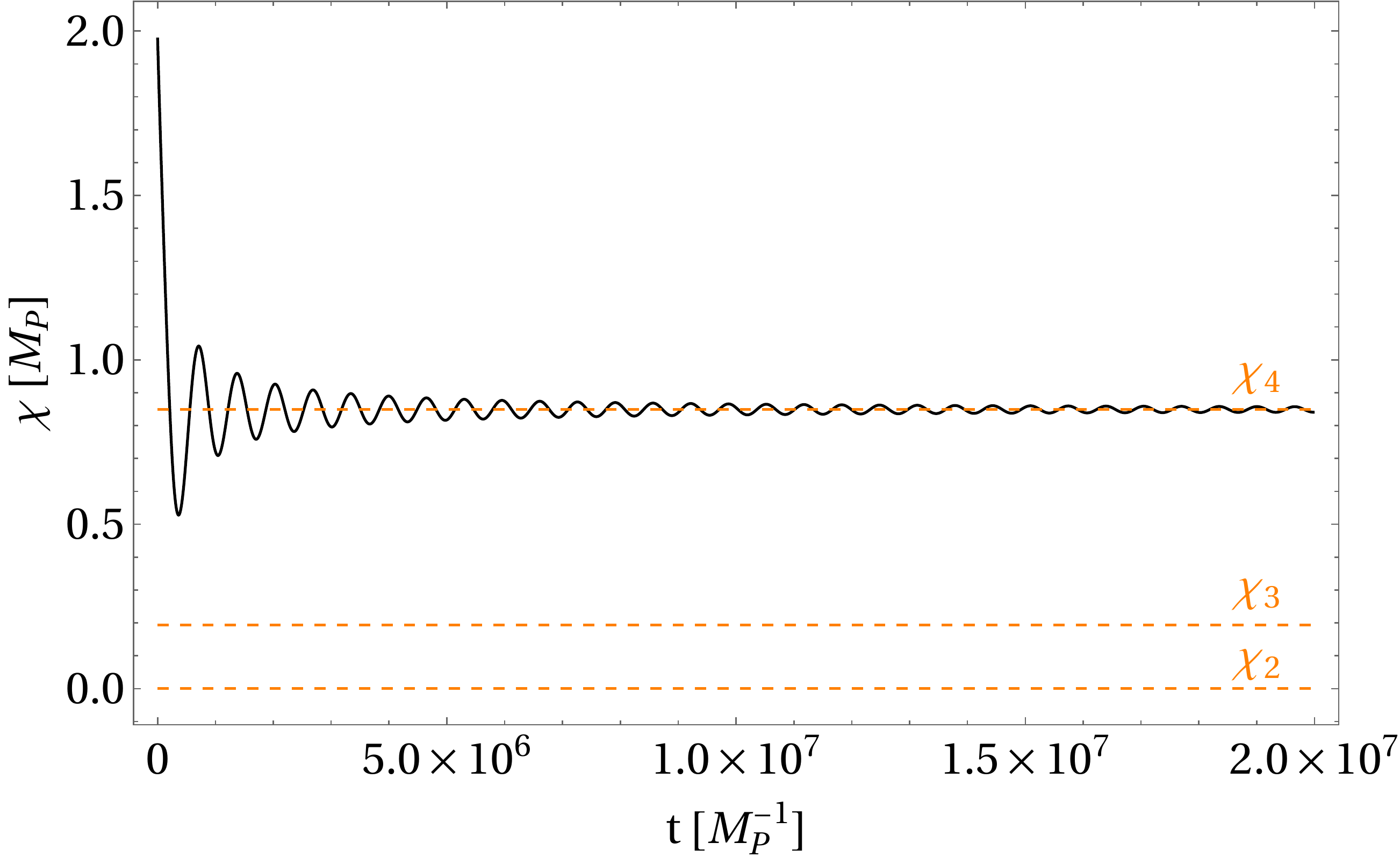}
\caption{The variation of the Einstein-frame inflaton field versus time in the symmetry-breaking model. In the figure, we have also specified the potential minima at $\chi_2$ and $\chi_4$, and its right local maximum at $\chi_3$ by orange dashed lines.}
\label{figure:chi-t-1}
\end{figure}

\begin{figure}
\centering
\includegraphics[scale=0.4]{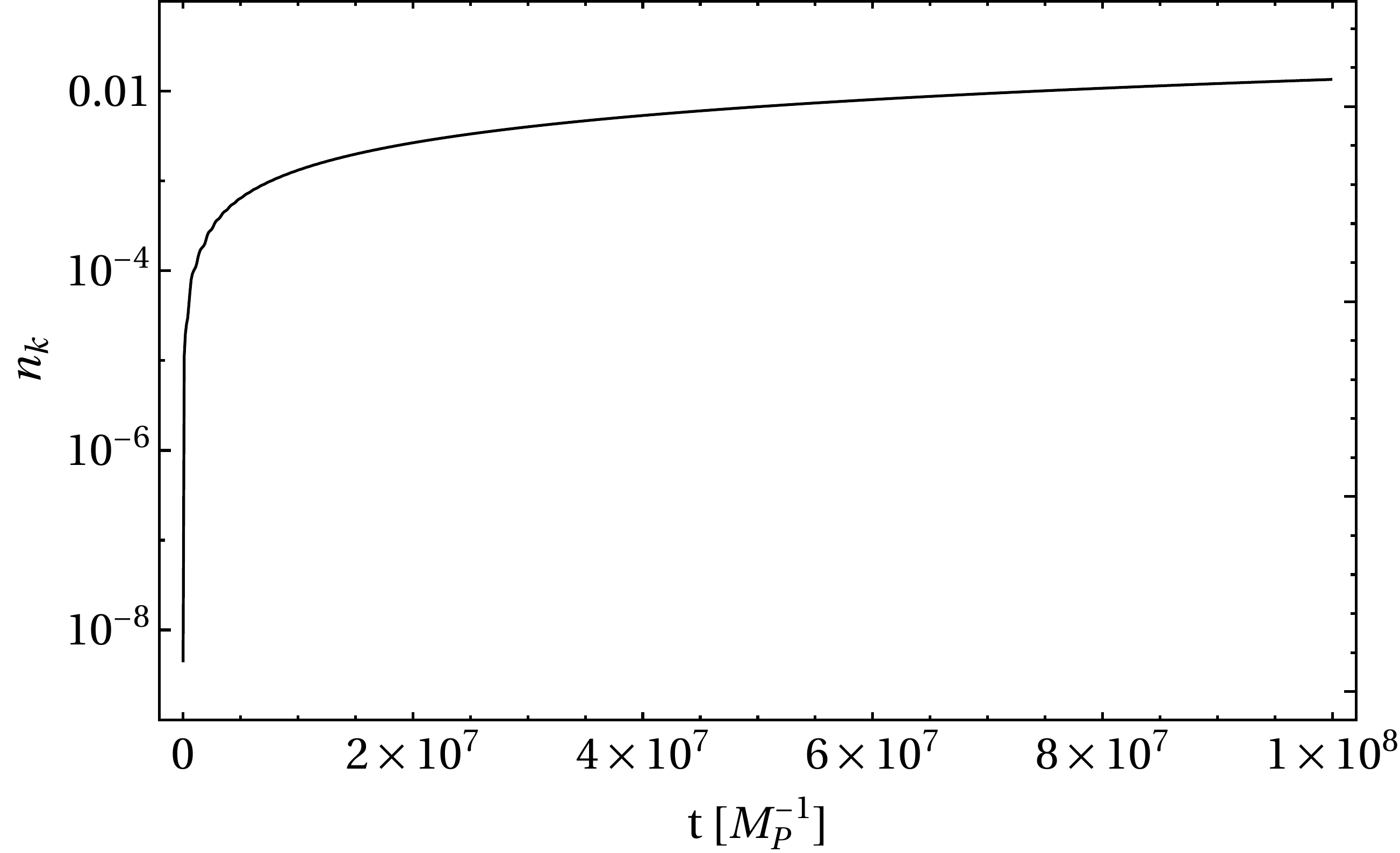}
\caption{The time variation of number density in the symmetry-breaking model.}
\label{figure:nk-1}
\end{figure}

The Jordan-frame potential \eqref{V0} for this case turns into
\begin{equation}
\label{V0-Case1}
V_{0}(\phi)=\frac{\text{\ensuremath{\lambda}}_{\mathrm{eff}}}{4}\phi^{2}\left(\phi-\mu\right)^{2} \, ,
\end{equation}
where we have defined $\mu\equiv\sqrt{2/\lambda_{\mathrm{eff}}}\,m$. This potential has a Mexican-hat shape that can be used to describe the spontaneous symmetry breaking in the early Universe \cite{liddle_lyth_2000}. Inserting this into Eq. \eqref{U}, we find the Einstein-frame potential as
\begin{equation}
\label{U-Case1}
U(\chi)=\frac{\text{\ensuremath{\text{\ensuremath{\lambda}}_{\mathrm{eff}}}}\phi^{2}(\chi)\left[\phi(\chi)-\mu\right]^{2}}{4\left[\xi\phi^{2}(\chi)+1\right]^{2}} \, .
\end{equation}
Here, we consider the model parameters as $\mu=0.01M_{P}$ and $\xi=10^{4}$. With these values, the diagram of the above potential has been depicted versus $\chi$ in Fig. \eqref{figure:U-1}. As we see in the figure, the potential has two minima in its shape which are located at
\begin{align}
\chi_{2} & =f(0)=0 \, ,
\label{chi2}
\\
\chi_{4} & =f(\mu)=\sqrt{\frac{6\xi+1}{\xi}}\,\sinh^{-1}\left[\mu\sqrt{\xi\left(6\xi+1\right)}\right]-\sqrt{6}\,\tanh^{-1}\left[\frac{\sqrt{6}\,\xi\mu}{\sqrt{\xi(6\xi+1)\mu^{2}+1}}\right] \, .
\label{chi4}
\end{align}
These minima are referred to as the super-symmetry and symmetry-breaking minima, respectively. In addition, the potential possesses two local maxima at $\chi_1$ and $\chi_3$ that can be determined by setting $U'(\chi) = 0$. Here, we focus on the inflationary scenario that takes place on the right branch of the symmetry-breaking minimum ($\chi_4$). Fixing the amplitude of the scalar perturbations at the horizon crossing as $\mathcal{P}_{s}\approx2.1\times10^{-9}$ according to the Planck 2018 constraints \cite{Planck:2018jri}, the self-interacting coupling constant is determined as $\text{\ensuremath{\lambda}}_{\mathrm{eff}}=0.1113$. With this value, the number of the simulations branes is obtained as $N\approx4$ which is remarkably less than the results of the M-flation setup with $N\sim10^{4}-10^{5}$. For the scalar spectral index and tensor-to-scalar ratio, we find respectively $n_{s}=0.9700$ and $r=0.0083$ which are in agreement with the Planck 2018 observational data \cite{Planck:2018jri}.

Now, we focus on the preheating process in this model, and for this purpose, we take $\omega=-3$. From this value of $\omega$, we deduce that we deal with the $\alpha$-mode with $j = 1$ in this model. In the preheating period after inflation, the evolution of the Einstein-frame scalar field in terms of cosmic time is obtained as depicted in Fig. \ref{figure:chi-t-1}. The figure shows that in this model, the $\chi$ field oscillates around the symmetry-breaking minimum, $\chi_4$, and it cannot cross the bump at the local maximum $\chi_3$, at all. $\chi$ never reaches the super-symmetry minimum located at $\chi_1$, therefore the tachyonic instability fails to happen in this case. We have illustrated the variation of the number density \eqref{nk} of this case with time in Fig. \ref{figure:nk-1}. The figure implies that the number density of this model remains always much less than unity and it cannot reach very large values at all. Thus, the parametric resonance cannot be accomplished effectively in this model to describe the particle production after inflation.


\section*{Acknowledgement}

We thank the referee for his/her valuable comments. We thank G. Felder for sending us a version of the LATTICEEASY code, which can be used to calculate the spectrum of GWs from preheating. This project has received funding/support from the European Union’s Horizon 2020 research and innovation programme under the Marie Sk\l{}odowska-Curie grant agreement No 860881-HIDDeN.


\bibliographystyle{fullsort.bst}
\bibliography{bibliography}

\providecommand{\href}[2]{#2}\begingroup\raggedright\begin{thebibliography}{10}

\bibitem{Linde:1990flp}
A.~D. Linde, {\em {Particle physics and inflationary cosmology}}, vol.~5.
\newblock 1990.

\bibitem{Traschen:1990sw}
J.~H. Traschen and R.~H. Brandenberger, ``{Particle Production During
  Out-of-equilibrium Phase Transitions},'' {\em Phys. Rev. D} {\bf 42} (1990)
  2491--2504.

\bibitem{Shtanov:1994ce}
Y.~Shtanov, J.~H. Traschen, and R.~H. Brandenberger, ``{Universe reheating
  after inflation},'' {\em Phys. Rev. D} {\bf 51} (1995) 5438--5455,
  \href{http://www.arXiv.org/abs/hep-ph/9407247}{{\tt hep-ph/9407247}}.

\bibitem{Kofman:1994rk}
L.~Kofman, A.~D. Linde, and A.~A. Starobinsky, ``{Reheating after inflation},''
  {\em Phys. Rev. Lett.} {\bf 73} (1994) 3195--3198,
  \href{http://www.arXiv.org/abs/hep-th/9405187}{{\tt hep-th/9405187}}.

\bibitem{Kofman:1997yn}
L.~Kofman, A.~D. Linde, and A.~A. Starobinsky, ``{Towards the theory of
  reheating after inflation},'' {\em Phys. Rev. D} {\bf 56} (1997) 3258--3295,
  \href{http://www.arXiv.org/abs/hep-ph/9704452}{{\tt hep-ph/9704452}}.

\bibitem{Ashoorioon:2009wa}
A.~Ashoorioon, H.~Firouzjahi, and M.~M. Sheikh-Jabbari, ``{M-flation: Inflation
  From Matrix Valued Scalar Fields},'' {\em JCAP} {\bf 06} (2009) 018,
  \href{http://www.arXiv.org/abs/0903.1481}{{\tt 0903.1481}}.

\bibitem{Ashoorioon:2011ki}
A.~Ashoorioon and M.~M. Sheikh-Jabbari, ``{Gauged M-flation, its UV sensitivity
  and Spectator Species},'' {\em JCAP} {\bf 06} (2011) 014,
  \href{http://www.arXiv.org/abs/1101.0048}{{\tt 1101.0048}}.

\bibitem{Ashoorioon:2009sr}
A.~Ashoorioon, H.~Firouzjahi, and M.~M. Sheikh-Jabbari, ``{Matrix Inflation and
  the Landscape of its Potential},'' {\em JCAP} {\bf 05} (2010) 002,
  \href{http://www.arXiv.org/abs/0911.4284}{{\tt 0911.4284}}.

\bibitem{Ashoorioon:2013oha}
A.~Ashoorioon, B.~Fung, R.~B. Mann, M.~Oltean, and M.~M. Sheikh-Jabbari,
  ``{Gravitational Waves from Preheating in M-flation},'' {\em JCAP} {\bf 03}
  (2014) 020, \href{http://www.arXiv.org/abs/1312.2284}{{\tt 1312.2284}}.

\bibitem{Ashoorioon:2014jja}
A.~Ashoorioon and M.~M. Sheikh-Jabbari, ``{Gauged M-flation After BICEP2},''
  {\em Phys. Lett. B} {\bf 739} (2014) 391--399,
  \href{http://www.arXiv.org/abs/1405.1685}{{\tt 1405.1685}}.

\bibitem{Linde:1983gd}
A.~D. Linde, ``{Chaotic Inflation},'' {\em Phys. Lett. B} {\bf 129} (1983)
  177--181.

\bibitem{Ooguri:2018wrx}
H.~Ooguri, E.~Palti, G.~Shiu, and C.~Vafa, ``{Distance and de Sitter
  Conjectures on the Swampland},'' {\em Phys. Lett. B} {\bf 788} (2019)
  180--184, \href{http://www.arXiv.org/abs/1810.05506}{{\tt 1810.05506}}.

\bibitem{Agrawal:2018own}
P.~Agrawal, G.~Obied, P.~J. Steinhardt, and C.~Vafa, ``{On the Cosmological
  Implications of the String Swampland},'' {\em Phys. Lett. B} {\bf 784} (2018)
  271--276, \href{http://www.arXiv.org/abs/1806.09718}{{\tt 1806.09718}}.

\bibitem{Planck:2018jri}
{\bf Planck} Collaboration, Y.~Akrami {\em et al.}, ``{Planck 2018 results. X.
  Constraints on inflation},'' {\em Astron. Astrophys.} {\bf 641} (2020) A10,
  \href{http://www.arXiv.org/abs/1807.06211}{{\tt 1807.06211}}.

\bibitem{Ashoorioon:2019kcy}
A.~Ashoorioon and K.~Rezazadeh, ``{Non-Minimal M-flation},'' {\em JHEP} {\bf
  07} (2020) 244, \href{http://www.arXiv.org/abs/1909.09806}{{\tt 1909.09806}}.

\bibitem{Ashoorioon:2011aa}
A.~Ashoorioon, U.~Danielsson, and M.~M. Sheikh-Jabbari, ``{1/N resolution to
  inflationary \ensuremath{\eta}-problem},'' {\em Phys. Lett. B} {\bf 713}
  (2012) 353--357, \href{http://www.arXiv.org/abs/1112.2272}{{\tt 1112.2272}}.

\bibitem{Kachru:2003sx}
S.~Kachru, R.~Kallosh, A.~D. Linde, J.~M. Maldacena, L.~P. McAllister, and
  S.~P. Trivedi, ``{Towards inflation in string theory},'' {\em JCAP} {\bf 10}
  (2003) 013, \href{http://www.arXiv.org/abs/hep-th/0308055}{{\tt
  hep-th/0308055}}.

\bibitem{Bezrukov:2007ep}
F.~L. Bezrukov and M.~Shaposhnikov, ``{The Standard Model Higgs boson as the
  inflaton},'' {\em Phys. Lett. B} {\bf 659} (2008) 703--706,
  \href{http://www.arXiv.org/abs/0710.3755}{{\tt 0710.3755}}.

\bibitem{Kofman:1995fi}
L.~Kofman, A.~D. Linde, and A.~A. Starobinsky, ``{Nonthermal phase transitions
  after inflation},'' {\em Phys. Rev. Lett.} {\bf 76} (1996) 1011--1014,
  \href{http://www.arXiv.org/abs/hep-th/9510119}{{\tt hep-th/9510119}}.

\bibitem{Felder:2000sf}
G.~N. Felder, L.~Kofman, A.~D. Linde, and I.~Tkachev, ``{Inflation after
  preheating},'' {\em JHEP} {\bf 08} (2000) 010,
  \href{http://www.arXiv.org/abs/hep-ph/0004024}{{\tt hep-ph/0004024}}.

\bibitem{Lyth:1995hj}
D.~H. Lyth and E.~D. Stewart, ``{Cosmology with a TeV mass GUT Higgs},'' {\em
  Phys. Rev. Lett.} {\bf 75} (1995) 201--204,
  \href{http://www.arXiv.org/abs/hep-ph/9502417}{{\tt hep-ph/9502417}}.

\bibitem{Adshead:2018oaa}
P.~Adshead, L.~Pearce, M.~Peloso, M.~A. Roberts, and L.~Sorbo, ``{Phenomenology
  of fermion production during axion inflation},'' {\em JCAP} {\bf 06} (2018)
  020, \href{http://www.arXiv.org/abs/1803.04501}{{\tt 1803.04501}}.

\bibitem{Adshead:2012kp}
P.~Adshead and M.~Wyman, ``{Chromo-Natural Inflation: Natural inflation on a
  steep potential with classical non-Abelian gauge fields},'' {\em Phys. Rev.
  Lett.} {\bf 108} (2012) 261302,
  \href{http://www.arXiv.org/abs/1202.2366}{{\tt 1202.2366}}.

\bibitem{Khlebnikov:1997di}
S.~Y. Khlebnikov and I.~I. Tkachev, ``{Relic gravitational waves produced after
  preheating},'' {\em Phys. Rev. D} {\bf 56} (1997) 653--660,
  \href{http://www.arXiv.org/abs/hep-ph/9701423}{{\tt hep-ph/9701423}}.

\bibitem{Easther:2006vd}
R.~Easther, J.~T. Giblin, Jr., and E.~A. Lim, ``{Gravitational Wave Production
  At The End Of Inflation},'' {\em Phys. Rev. Lett.} {\bf 99} (2007) 221301,
  \href{http://www.arXiv.org/abs/astro-ph/0612294}{{\tt astro-ph/0612294}}.

\bibitem{Dufaux:2007pt}
J.~F. Dufaux, A.~Bergman, G.~N. Felder, L.~Kofman, and J.-P. Uzan, ``{Theory
  and Numerics of Gravitational Waves from Preheating after Inflation},'' {\em
  Phys. Rev. D} {\bf 76} (2007) 123517,
  \href{http://www.arXiv.org/abs/0707.0875}{{\tt 0707.0875}}.

\bibitem{Adshead:2018doq}
P.~Adshead, J.~T. Giblin, and Z.~J. Weiner, ``{Gravitational waves from gauge
  preheating},'' {\em Phys. Rev. D} {\bf 98} (2018), no.~4, 043525,
  \href{http://www.arXiv.org/abs/1805.04550}{{\tt 1805.04550}}.

\bibitem{Adshead:2019lbr}
P.~Adshead, J.~T. Giblin, M.~Pieroni, and Z.~J. Weiner, ``{Constraining axion
  inflation with gravitational waves from preheating},'' {\em Phys. Rev. D}
  {\bf 101} (2020), no.~8, 083534,
  \href{http://www.arXiv.org/abs/1909.12842}{{\tt 1909.12842}}.

\bibitem{Cui:2021are}
Y.~Cui and E.~I. Sfakianakis, ``{Detectable gravitational wave signals from
  inflationary preheating},'' {\em Phys. Lett. B} {\bf 840} (2023) 137825,
  \href{http://www.arXiv.org/abs/2112.00762}{{\tt 2112.00762}}.

\bibitem{Lozanov:2019ylm}
K.~D. Lozanov and M.~A. Amin, ``{Gravitational perturbations from oscillons and
  transients after inflation},'' {\em Phys. Rev. D} {\bf 99} (2019), no.~12,
  123504, \href{http://www.arXiv.org/abs/1902.06736}{{\tt 1902.06736}}.

\bibitem{Antusch:2016con}
S.~Antusch, F.~Cefala, and S.~Orani, ``{Gravitational waves from oscillons
  after inflation},'' {\em Phys. Rev. Lett.} {\bf 118} (2017), no.~1, 011303,
  \href{http://www.arXiv.org/abs/1607.01314}{{\tt 1607.01314}}. [Erratum:
  Phys.Rev.Lett. 120, 219901 (2018)].

\bibitem{Amin:2018xfe}
M.~A. Amin, J.~Braden, E.~J. Copeland, J.~T. Giblin, C.~Solorio, Z.~J. Weiner,
  and S.-Y. Zhou, ``{Gravitational waves from asymmetric oscillon dynamics?},''
  {\em Phys. Rev. D} {\bf 98} (2018) 024040,
  \href{http://www.arXiv.org/abs/1803.08047}{{\tt 1803.08047}}.

\bibitem{Hiramatsu:2020obh}
T.~Hiramatsu, E.~I. Sfakianakis, and M.~Yamaguchi, ``{Gravitational wave
  spectra from oscillon formation after inflation},'' {\em JHEP} {\bf 03}
  (2021) 021, \href{http://www.arXiv.org/abs/2011.12201}{{\tt 2011.12201}}.

\bibitem{Kou:2021bij}
X.-X. Kou, J.~B. Mertens, C.~Tian, and S.-Y. Zhou, ``{Gravitational waves from
  fully general relativistic oscillon preheating},'' {\em Phys. Rev. D} {\bf
  105} (2022), no.~12, 123505, \href{http://www.arXiv.org/abs/2112.07626}{{\tt
  2112.07626}}.

\bibitem{Garcia-Bellido:2007nns}
J.~Garcia-Bellido and D.~G. Figueroa, ``{A stochastic background of
  gravitational waves from hybrid preheating},'' {\em Phys. Rev. Lett.} {\bf
  98} (2007) 061302, \href{http://www.arXiv.org/abs/astro-ph/0701014}{{\tt
  astro-ph/0701014}}.

\bibitem{Garcia-Bellido:2007fiu}
J.~Garcia-Bellido, D.~G. Figueroa, and A.~Sastre, ``{A Gravitational Wave
  Background from Reheating after Hybrid Inflation},'' {\em Phys. Rev. D} {\bf
  77} (2008) 043517, \href{http://www.arXiv.org/abs/0707.0839}{{\tt
  0707.0839}}.

\bibitem{Dufaux:2010cf}
J.-F. Dufaux, D.~G. Figueroa, and J.~Garcia-Bellido, ``{Gravitational Waves
  from Abelian Gauge Fields and Cosmic Strings at Preheating},'' {\em Phys.
  Rev. D} {\bf 82} (2010) 083518,
  \href{http://www.arXiv.org/abs/1006.0217}{{\tt 1006.0217}}.

\bibitem{Ashoorioon:2022raz}
A.~Ashoorioon, K.~Rezazadeh, and A.~Rostami, ``{NANOGrav Signal from the End of
  Inflation and the LIGO Mass and Heavier Primordial Black Holes},''
  \href{http://www.arXiv.org/abs/2202.01131}{{\tt 2202.01131}}.

\bibitem{Berlin:2021txa}
A.~Berlin, D.~Blas, R.~Tito~D'Agnolo, S.~A.~R. Ellis, R.~Harnik, Y.~Kahn, and
  J.~Sch\"utte-Engel, ``{Detecting high-frequency gravitational waves with
  microwave cavities},'' {\em Phys. Rev. D} {\bf 105} (2022), no.~11, 116011,
  \href{http://www.arXiv.org/abs/2112.11465}{{\tt 2112.11465}}.

\bibitem{Felder:2000hq}
G.~N. Felder and I.~Tkachev, ``{LATTICEEASY: A Program for lattice simulations
  of scalar fields in an expanding universe},'' {\em Comput. Phys. Commun.}
  {\bf 178} (2008) 929--932,
  \href{http://www.arXiv.org/abs/hep-ph/0011159}{{\tt hep-ph/0011159}}.

\bibitem{ADMX:2021nhd}
{\bf ADMX} Collaboration, C.~Bartram {\em et al.}, ``{Search for Invisible
  Axion Dark Matter in the 3.3\textendash{}4.2\,\,\ensuremath{\mu}eV Mass
  Range},'' {\em Phys. Rev. Lett.} {\bf 127} (2021), no.~26, 261803,
  \href{http://www.arXiv.org/abs/2110.06096}{{\tt 2110.06096}}.

\bibitem{HAYSTAC:2018rwy}
{\bf HAYSTAC} Collaboration, L.~Zhong {\em et al.}, ``{Results from phase 1 of
  the HAYSTAC microwave cavity axion experiment},'' {\em Phys. Rev. D} {\bf 97}
  (2018), no.~9, 092001, \href{http://www.arXiv.org/abs/1803.03690}{{\tt
  1803.03690}}.

\bibitem{McAllister:2017lkb}
B.~T. McAllister, G.~Flower, J.~Kruger, E.~N. Ivanov, M.~Goryachev,
  J.~Bourhill, and M.~E. Tobar, ``{The ORGAN Experiment: An axion haloscope
  above 15 GHz},'' {\em Phys. Dark Univ.} {\bf 18} (2017) 67--72,
  \href{http://www.arXiv.org/abs/1706.00209}{{\tt 1706.00209}}.

\bibitem{CAPP:2020utb}
{\bf CAPP} Collaboration, O.~Kwon {\em et al.}, ``{First Results from an Axion
  Haloscope at CAPP around 10.7 $\mu$eV},'' {\em Phys. Rev. Lett.} {\bf 126}
  (2021), no.~19, 191802, \href{http://www.arXiv.org/abs/2012.10764}{{\tt
  2012.10764}}.

\bibitem{liddle_lyth_2000}
A.~R. Liddle and D.~H. Lyth, {\em Cosmological Inflation and Large-Scale
  Structure}.
\newblock Cambridge University Press, 2000.

\end{thebibliography}\endgroup


\end{document}